\documentclass[%
superscriptaddress,
 amsmath,amssymb,
 aps,
 prc,
 twocolumn
]{revtex4-2}

\usepackage{graphicx}
\usepackage{dcolumn}
\usepackage{bm}
\usepackage{multirow}
\usepackage{subfiles} 
\usepackage[usenames,dvipsnames]{xcolor}
\definecolor{airforceblue}{rgb}{0.36, 0.54, 0.66}

\usepackage[colorlinks,bookmarks=false,
hypertexnames=true,allcolors=airforceblue]{hyperref}
\usepackage[modulo]{lineno}

\begin{document}


\title{Experimental Study of the $^{38}$S Excited Level Scheme}

\author{C.~R.~Hoffman}
\email{calem.hoffman@gmail.com}
\affiliation{Physics Division, Argonne National Laboratory, 9700 S. Cass Ave., Argonne, IL 60439, USA}%
\author{R.~S.~Lubna}
\affiliation{TRIUMF, 4004 Wesbrook Mall, Vancouver, British Columbia V6T 2A3, Canada}%
\affiliation{Facility for Rare Isotope Beams, Michigan State University, East Lansing, Michigan 48824, USA}%
\author{E.~Rubino}
\altaffiliation[Present Address: ]{Lawrence Livermore National Laboratory, Livermore, CA 94550, USA.}
\affiliation{Department of Physics, Florida State University, Tallahassee, Florida 32306, USA}%
\author{S.~L.~Tabor}
\affiliation{Department of Physics, Florida State University, Tallahassee, Florida 32306, USA}%
\author{K.~Auranen}
\altaffiliation[Present Address: ]{Accelerator Laboratory, Department of Physics, University of Jyv\"{a}skyl\"{a}, FI-40014 Jyv\"{a}skyl\"{a}, Finland.}
\affiliation{Physics Division, Argonne National Laboratory, 9700 S. Cass Ave., Argonne, IL 60439, USA}%
\author{P.~C.~Bender}
\affiliation{Department of Physics and Applied Physics, University of Massachusetts Lowell, Lowell, Massachusetts 01854, USA}%
\author{C.~M.~Campbell}
\affiliation{Nuclear Science Division, Lawrence Berkeley National Laboratory, Berkeley, California 94720, USA}%
\author{M.~P.~Carpenter}
\affiliation{Physics Division, Argonne National Laboratory, 9700 S. Cass Ave., Argonne, IL 60439, USA}%
\author{J.~Chen}
\affiliation{Physics Division, Argonne National Laboratory, 9700 S. Cass Ave., Argonne, IL 60439, USA}%
\author{M.~Gott}
\altaffiliation[Present Address: ]{Oak Ridge National Laboratory, Oak Ridge, Tennessee 37831, USA.}
\affiliation{Physics Division, Argonne National Laboratory, 9700 S. Cass Ave., Argonne, IL 60439, USA}%
\author{J.~P.~Greene}
\affiliation{Physics Division, Argonne National Laboratory, 9700 S. Cass Ave., Argonne, IL 60439, USA}%
\author{D.~E.~M.~Hoff}
\altaffiliation[Present Address: ]{Lawrence Livermore National Laboratory, Livermore, CA 94550, USA.}
\affiliation{Department of Physics and Applied Physics, University of Massachusetts Lowell, Lowell, Massachusetts 01854, USA}%
\author{T.~Huang}
\affiliation{Physics Division, Argonne National Laboratory, 9700 S. Cass Ave., Argonne, IL 60439, USA}
\affiliation{Institute of Modern Physics, Chinese Academy of Sciences, Lanzhou 730000, China}%
\author{H.~Iwasaki}
\affiliation{Facility for Rare Isotope Beams, Michigan State University, East Lansing, Michigan 48824, USA}%
\affiliation{Department of Physics and Astronomy, Michigan State University, East Lansing, MI 48824, USA}%
\author{F.~G.~Kondev}
\affiliation{Physics Division, Argonne National Laboratory, 9700 S. Cass Ave., Argonne, IL 60439, USA}%
\author{T.~Lauritsen}
\affiliation{Physics Division, Argonne National Laboratory, 9700 S. Cass Ave., Argonne, IL 60439, USA}%
\author{B.~Longfellow}
\affiliation{Lawrence Livermore National Laboratory, Livermore, CA 94550, USA}
\author{C.~Santamaria}
\altaffiliation[Present Address: ]{Morgan State University, 1700 East Cold Spring Lane, Baltimore, Maryland 21251, USA.}
\affiliation{Nuclear Science Division, Lawrence Berkeley National Laboratory, Berkeley, California 94720, USA}%
\author{D.~Seweryniak}
\affiliation{Physics Division, Argonne National Laboratory, 9700 S. Cass Ave., Argonne, IL 60439, USA}%
\author{T.~L.~Tang}
\altaffiliation[Present Address: ]{Florida State University, Tallahassee, Florida 32306, USA}
\affiliation{Physics Division, Argonne National Laboratory, 9700 S. Cass Ave., Argonne, IL 60439, USA}%
\author{G.~L.~Wilson}
\affiliation{Department of Physics and Astronomy, Louisiana State University, Baton Rouge, Louisiana 70803, USA}%
\affiliation{Physics Division, Argonne National Laboratory, 9700 S. Cass Ave., Argonne, IL 60439, USA}%
\author{J.~Wu}
\altaffiliation[Present Address: ]{Brookhaven National Laboratory, National Nuclear Data Center, Upton, New York 11973, USA.}
\affiliation{Physics Division, Argonne National Laboratory, 9700 S. Cass Ave., Argonne, IL 60439, USA}%
\author{S.~Zhu}
\altaffiliation[]{Deceased.}
\affiliation{Physics Division, Argonne National Laboratory, 9700 S. Cass Ave., Argonne, IL 60439, USA}%

\date{\today}
\begin{abstract}
Information on the $^{38}$S level scheme was expanded through experimental work utilizing a fusion-evaporation reaction and in-beam $\gamma$-ray spectroscopy. Prompt $\gamma$-ray transitions were detected by the Gamma-Ray Energy Tracking Array (GRETINA) and recoiling $^{38}$S residues were selected by the Fragment Mass Analayzer (FMA). Tools based on machine-learning techniques were developed and deployed for the first time in order to enhance the unique selection of $^{38}$S residues and identify any associated $\gamma$-ray transitions. The new level information, including the extension of the even-spin yrast sequence through $J^{\pi} = 8^{(+)}$, was interpreted in terms of a basic single-particle picture as well shell-model calculations which incorporated the empirically derived FSU interaction. A comparison between the properties of the yrast states in the even-$Z$ $N=22$ isotones from $Z=14$ to $20$, and for $^{36}$Si-$^{38}$S in particular, was also presented with an emphasis on the role and influence of the neutron $1p_{3/2}$ orbital on the structure in the region.
\end{abstract}

\maketitle

\section{\label{sec:intro}Introduction \& Background}

First discovered in 1958~\cite{ref:Net58}, $^{38}$S is comprised of sixteen protons ($Z=16$), twenty-two neutrons ($N=22$), and isospin $T = 3$. Its positioning on the chart of the nuclides is such that two valence neutrons reside outside of the traditional $N=20$ shell closure defined by the $1s0d$-$0f1p$ shell gap while valence protons fill two-thirds of the $1s0d$ shell including a large fraction of the $\pi0d_{5/2}$ orbital. The ground state and low-lying structure in $^{38}$S lends itself to a competition between the single-particle and coherent deformation-driven aspects of the $\pi1s0d$ and $\nu0f1p$ orbitals. In particular, the proximity and occupancy of the proton $0d_{5/2}-1s_{1/2}-0d_{3/2}$ orbitals and the absence of a robust $N=28$ shell-gap between the $\nu0f_{7/2}-1p_{3/2}$ orbitals, provide scenarios for strong proton-neutron quadrupole-based correlations to thrive. The higher-lying excited levels in $^{38}$S are expected to be additionally influenced by various particle-hole $N=20$ cross-shell excitations. Similar competitions and effects are known to drive the deformation that has been observed in the low-lying levels of the $N\approx28$ Si and S isotopes (see Sec.~4.4 of Ref.~\cite{ref:Nowacki2021} and references therein).

The $^{38}$S ground state has a lifetime of T$_{1/2}=170.3(7)$~min and $\beta^-$ decays with a branch of 100\%~\cite{ref:ensdf}. 
A number of excited levels have been observed below 7~MeV based on data collected from $\beta$-delayed $\gamma$-ray spectroscopy~\cite{ref:Duf86}, two-particle transfer reactions - some that incorporated $\gamma$-ray detection~\cite{ref:Fif84,ref:May84,ref:Dav85,ref:Oln86,ref:War87}, and in-beam $\gamma$-ray spectroscopy measurements utilizing deep inelastic reactions~\cite{ref:For94,ref:Oll04,ref:Wang2010,ref:Wang} or intermediate energy fragmentation~\cite{ref:Lun16}. 
A summary of the level energies and spin-parity ($J^{\pi}$) values was presented in Fig.~6.46 of Ref.~\cite{ref:Wang}. 
Firm spin-parity assignments were established only for the yrast even-$J$ levels through $J^{\pi}=4^+$ and the $2^+_2$ level at 2.806~MeV. 
A probable candidate for the yrast $6^+_1$ level was first observed in the work of Ref.~\cite{ref:For94}. 
Candidates for possible negative parity (intruder) states starting at around 3.5~MeV up through $\sim$6~MeV were also identified, primarily in the ($t$,$p$) and $\beta$ decay works~\cite{ref:Duf86,ref:Dav85}. 
Additional spectroscopic information with respect to the excited levels in $^{38}$S, such as lifetimes and transitions stengths of various excited levels and the ratio of the multipole matrix elements ($M_n/M_p$) plus the $g$ factor for the $2^+_1$~$\rightarrow$~$0^+_1$ transition, has been extracted from in-beam Coulomb excitation, inelastic scattering, and deep-inelastic reaction data~\cite{ref:Sch96,ref:Kel97,ref:Suo97,ref:Stu06,ref:Dav06,ref:Ots07,ref:Longfellow2021,ref:Grocutt2022}.

The energies of the established yrast levels in $^{38}$S have been well described by having valence protons contained within the $1s0d$ shell and valence neutrons contained with the $0f1p$ shell. This was demonstrated by the weak-coupling model prescription of Bansal and French~\cite{ref:Bansal1964} presented in Fig.~4 of Ref.~\cite{ref:Dav85}, for example. Unsurprisingly, a number of effective interactions derived within this model space have also been successful in reproducing the observed level energies and the even-$J$ yrast states in particular, for instance, the results of the $SDPF$ interactions of Refs.~\cite{ref:Vanderpoel1982,ref:Vanderpoel1983,ref:War86} shown in Fig. 4 of both Refs.~\cite{ref:Dav85} and ~\cite{ref:Oln86}. More recently developed interactions such as the SDPF-MU interaction~\cite{ref:Uts01,ref:Utsuno2012}, the sdpf-nr interaction~\cite{ref:Nummela2001}, and the SDPF-U interaction~\cite{ref:Now09}, have also been successful in describing the low-lying level energies (see Fig.~4 of Ref.~\cite{ref:Lun16} and Fig.~6.46 of Ref.~\cite{ref:Wang2010}). Shell-model interactions which have been sure to incorporate the $1p_{3/2}$ neutron orbital within their model space, have shown promise in describing the degree and nature of the deformation in $^{38}$S. This includes agreement with experimental B($E2$) transition strengths~\cite{ref:Sch96,ref:Longfellow2021,ref:Grocutt2022} as well as the measured $0^+\rightarrow2^+_1$ $g$ factor~\cite{ref:Dav06,ref:Stu06}. A common theme building from these past works has been an emphasis on the key role played by the $\nu1p_{3/2}$ orbital in generating the proper amount of coherent proton-neutron (quadrupole) correlations in the low-lying levels.

The present work describes the first investigation of the $^{38}$S level scheme through population in a fusion-evaporation reaction (Section~\ref{sec:exp}). In doing so, a significant number of new excited levels and transitions have been determined from the in-beam $\gamma$-ray data, uniquely selected following $^{38}$S recoil identification (Section~\ref{sec:ana}). In particular, the yrast even-$J$ levels have been extended to $8^{(+)}$ and a number of higher-$J$ candidates have been found. Multipolarities have been deduced for some transitions based on $\gamma$-ray yields and were used to assign or suggest $J$-values where possible (Section~\ref{sec:res}). The resulting level scheme is discussed in terms of simple symmetry arguments, the nearby nuclei in the region, the role of the $\nu1p_{3/2}$ orbital, and comparisons are made directly with shell-model calculations using the FSU cross-shell interaction~\cite{ref:Lubna2020} (Section~\ref{sec:dis}).

\section{\label{sec:exp}Experimental Details}

Excited states in $^{38}$S were populated in the $^{18}$O($^{22}$Ne,2$p$) fusion-evaporation reaction. A 48.5~MeV $^{22}$Ne$^{6+}$ primary beam at an intensity of 30~particle-nano-Amperes was provided by the Argonne Tandem Linear Accelerator System (ATLAS) facility located at Argonne National Laboratory. The $^{18}$O targets were prepared by electrodeposition. A 1~mL aliquot of 99\% $^{18}$O H$_2$O was added to the deposition chamber seated on top of a $\sim$1.1~mg/cm$^{2}$ Ta foil. Electrodeposition was performed using a platinum cathode and the Ta foil as the anode, which was held at a constant $+100$~V DC for approximately 5 hours. As the $^{18}$O diffused into the Ta foil, the surface passivated resulting in the current dropping from several hundred $\mu$A to tens of $\mu$A. Mass gain and alpha loss measurements indicated a few hundred $\mu$g/cm$^2$ of $^{18}$O were diffused into one side of the Ta foil. The Ta foil was arranged so that the $^{18}$O material resided on the downstream target side allowing the best possible release of heavy-ion recoils. The beam energy was chosen to take into account the energy loss through $\sim1/2$ of the Ta foil and assumed that the $^{18}$O material was evenly distributed in the latter half of the foil. No degradation in the amount or distribution of the $^{18}$O material was observed over the duration of the run ($\sim$6 days).

Prompt $\gamma$-ray transitions emanating from nuclei following their population in the fusion-evaporation reaction were detected in the Gamma-Ray Energy Tracking Array (GRETINA)~\cite{ref:Lee04,ref:Pas13}. GRETINA consisted of 12 HPGe modules and was positioned to cover polar-angles between $70^{\circ}<\theta<170^{\circ}$ relative to the beam direction. Relative energy-dependent detection efficiencies and energy-response calibrations were carried out with standard $\gamma$-ray sources of $^{152}$Eu and $^{56}$Co. The data collected was processed in an add-back mode whereby $\gamma$-ray interactions that occurred within 10~cm of each other were energy summed. Both the source and in-beam $\gamma$-ray data was processed the same way. The position information from the largest-energy interaction were used to define the outgoing $\gamma$-ray angle for Doppler reconstruction. Similarly, the timestamp from largest-energy interaction was used as the timestamp for that event and used for relative-timing coincidences. An average recoil velocity of $\beta=0.003375(15)$/c was uniquely determined for the $^{38}$S recoils from a fit of the known 1293-keV and 1534-keV $\gamma$-ray energies as a function of polar angle. An energy width of $\sim$0.5\% FWHM was measured for the 1293-keV $\gamma$-ray in $^{38}$S.

The Fragment Mass Analayzer (FMA)~\cite{ref:Dav92} provided selection capabilities for recoiling $^{38}$S ions via energy-to-charge, $E/q$, and mass-to-charge, $A/q$, ratios. The magnetic and electric elements of the FMA were optimized for the transmission of $^{38}$S$^{8+}$ at 18.7~MeV, i.e., $A/q=38/8=4.75$ and $E/q=18.7/8\approx 2.338$. Movable slits in the dispersive (horizontal) direction were used at the FMA focal plane to suppress recoils with similar $A/q$ values and to reduce the intensity of scattered un-reacted primary beam. Both dispersive and non-dispersive (vertical) position information was collected on an event-by-event basis at the FMA focal plane by a parallel-grid avalanche counter (PGAC) filled with isobutane gas to a pressure of $\sim$400~Pa (3~Torr). Timing signals to both trigger data collection and for use in the event-by-event time-of-flight reconstruction originated from the PGAC anode signal. An ionization chamber (IC) having a segmented-anode readout consisting of two 5~cm long sections followed by a 20~cm long section, was positioned directly behind the PGAC. The IC was operated with isobutane gas at a pressure of $\sim$1067~Pa (8~Torr) which ensured the stopping of all recoils of interest within its volume. The recoil-ion energy information was recorded independently for each of the three sections, $E_i$ ($i=1-3$).

All data were collected by a digital data acquisition system which sampled the input analog signals at a frequency of 100~MHz. The valid condition to activate data collection was a relative time-coincidence of $<1\mu$s between timestamps of the PGAC anode and GRETINA - when the $\gamma$-ray multiplicity condition of $M_{\gamma}>0$ was met. Typical data-collection rates were on the order of $\sim$5~kHz.

\section{\label{sec:ana}Experimental \& Analysis Methods}

\subsection{$^{38}$S recoil selection}
\begin{figure}[htb]
    \centering
    \includegraphics[scale=0.21]{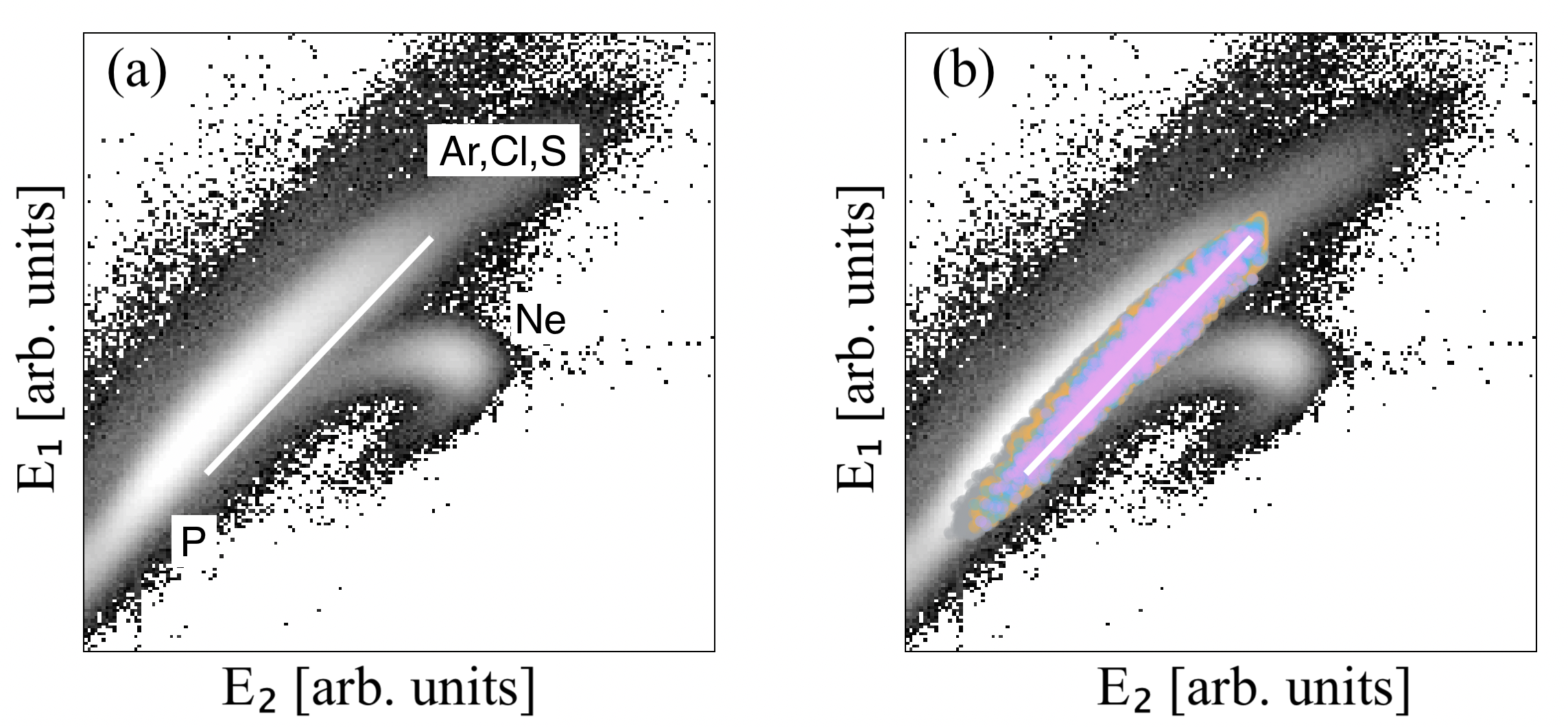}\\
    \includegraphics[scale=0.21]{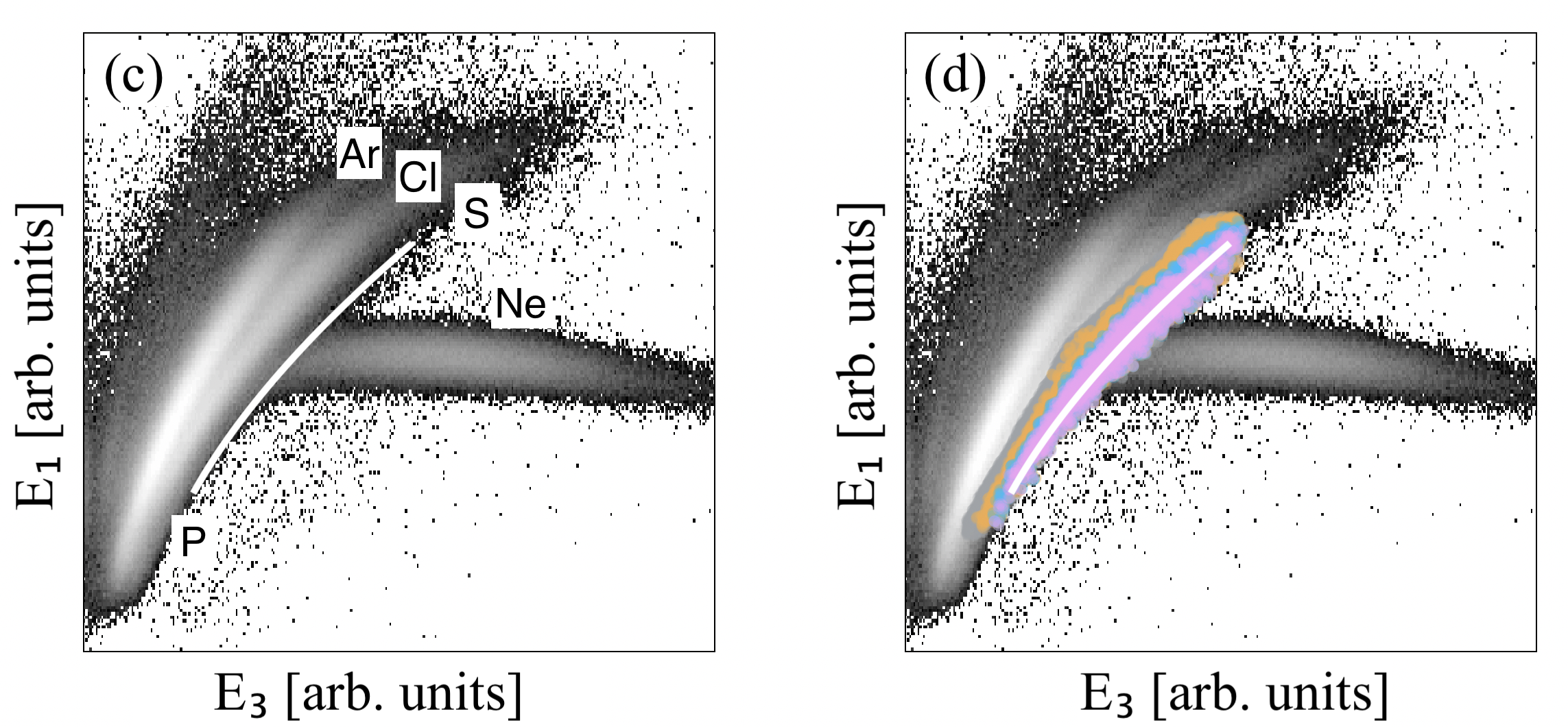}\\
    \includegraphics[scale=0.21]{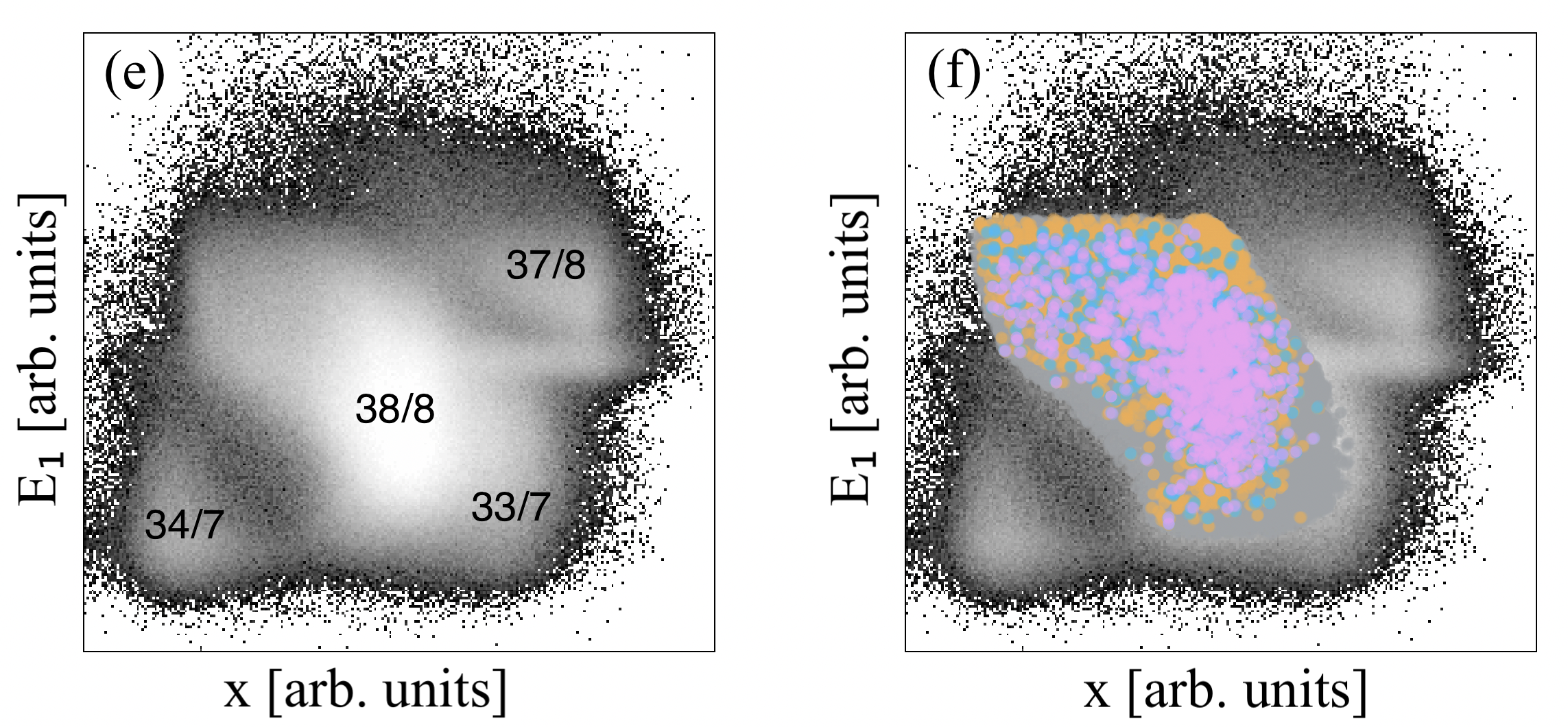}
    \caption{The energy of the first ion chamber section ($E_{1}$) plotted against the energy of the second section ($E_2$) (a)-(b), the third section ($E_3$) (c)-(d), and the dispersive position at the FMA focal plane ($x$) in (e)-(f). The time-coincidence gate $|\Delta T_{\gamma -PGAC}|<200$~ns was applied to the grey-scale histograms which are the same across each row. Element ($Z$) labels in (a) and (c) identify the general regions of interest, while the white line identifies the central region covered by $^{38}$S recoils. The labels in (e) correspond to $A/q$ values. The overlaid spectra in (b), (d), and (f) include data which define the manual cuts (grey points) and from differing model output values, $k_{ml}>0.25$ (yellow), 0.60 (blue), and 0.81 (pink), as defined in Fig.~\ref{fig:fig4} (see text for additional details). The white line in (a) - (d) represents the central region of the S isotopes and is the same for each row.}
    \label{fig:fig1}
\end{figure}

There are a number of example two-proton evaporation analyses prior to this work which involved the selection and identification of isotopes of interest by the FMA~\cite{ref:Fre04,ref:Zhu06,ref:Dea10}. Each of those leveraged some combination of the following quantities to extract the events of interest: i) the relative times between prompt $\gamma$-rays and the PGAC ($T_{\gamma -PGAC}$) to provide suppression of random coincidences, ii) the individual or summed ionization chamber energy information ($E_{i}, E_{ij}, E_{ijk}$) to identify the element ($Z$) of interest, iii), the dispersive position of the recoil at the FMA focal plane ($x$) to define $A/q$, and iv) the combined (total) ion chamber energy, $E_{123}$, with the recoil time-of-flight ($T_{\gamma -PGAC}$) to distinguish $A$ through the classical mass relation, $m\sim E_{123}T_{\gamma -PGAC}^2$. In the present work, the beam energy and extra target thickness due to the Ta foil matrix resulted in relatively low energies for the recoiling $^{38}$S ions ($\lesssim 20$~MeV). The consequence was a less than optimal identification of $^{38}$S recoils based on a series of manual selections or gates applied to the quantities listed above. For instance, as viewed in Figs.~\ref{fig:fig1}(a) - (d), the central region of the S recoils (indicated by the white line) from the ionization chamber energies overlapped with the neighboring isobar $^{38}$Cl. In addition, the dispersive position $x\propto A/q$ plot in Fig.~\ref{fig:fig1}(e) identifies an overlap between the $A/q=38/8$ recoils of interest with a nearby $A/q=33/7$ ambiguity belonging to $^{33}$P$^{7+}$.

\subsubsection{Application of a fully-connected feed-forward neural network (NN) model}
To improve upon the selection of $^{38}$S recoils over the application of manual gating alone, a fully-connected feed-forward neural network (NN) model was developed and implemented. This is the first application of such a model to be used for assisting with recoil selection at the FMA. The power of trainable NN models for various classifications scenarios is now wide-spread and impactful across nuclear science research, see for example Ref.~\cite{ref:Boehnlein2021} and references therein. In the present work, we developed and trained through supervision an NN model with the ability to process our event data in a goal oriented One-Vs-All mode. In this mode, the $^{38}$S recoil data is meant to be distinguished, or classified, against all other types of recoil data. Types of data not belonging to $^{38}$S recoils included other fusion-evaporation reaction channels resulting in $^{38}$Cl and $^{33}$P recoils, random data originating from the $^{181}$Ta lattice and scattered primary beam, and Compton-scattering background data. The framework of the NN model was designed to output a single value, $k_{ml}$, ranging from 0 -- 1, for each data input into the model. In one sense, $k_{ml}$ is representative of the degree or likely-hood to which the model has found the corresponding input data to be to classified as $^{38}$S (approaching a value of 1) or not (approaching towards 0). An individual $\gamma$-ray energy, after the add-back procedures described in Section~\ref{sec:exp}, was used to define a unique set of input data to the NN model. For each $\gamma$ ray, eleven experimental values were associated with it, defining the input size of the first NN model layer. For example, an event with a $\gamma$-ray multiplicity of 5, $M_{\gamma}=5$, was considered as five independent sets of data inputs into the model. The input values included the individual and summing permutations of the ion chamber energies ($E_1$,$E_2$,$E_3$,$E_{12}$,$E_{13}$,$E_{23}$,$E_{123}$), the FMA focal plane dispersive plane position information ($x$), the relative recoil time-of-flight from the target to the FMA focal plane ($T_{\gamma -PGAC}$), the calculated mass ($m\sim E_{123} T^{2}_{\gamma -PGAC}$), and the $\gamma$-ray multiplicity ($M_{\gamma}$) of the complete event to which the individual $\gamma$-ray belonged.  The number of input parameters could have been reduced due to redundancies and correlations, for example, the parameters used to generate $m$ were all included independently. However, the calculated values were already readily available and the number of inputs were relatively small. As is standard practice, each input value range was independently normalized to span from 0 -- 1 to reduce the possibility of any biases appearing throughout the NN model training.

The framework of the NN model itself was standard and considered to be shallow. It consisted of three total layers, an input layer with dimensions (x11,x40), a hidden layer (x40,x20), and an output layer (x20,x1). Each corresponding layer was linearly- and fully-connected with the previous one. The rectified linear unit (Relu) and Parametric Relu (PRelu) functions~\cite{ref:He2015} were applied to the input and hidden layers, respectively. A Sigmoid activation function was applied to the output layer. A dropout function~\cite{ref:Srivastava2014} was active between the input and hidden layers with a 20\% probability for zeroing any individual neuron within that layer. In total, there were 1322 trainable parameters: 480 for the input layer, 820+1 for the hidden layer plus PRelu function, and 21 for the output layer. The NN model was trained under supervision through gradient descent and back-propagation using the Binary Cross Entropy Loss (BCELoss) function and Adam optimizer~\cite{ref:Kingma2017} with a learning rate of $10^{-3}$ and exponential decay rate values of $\beta_1=0.9$ and $\beta_2=0.999$ on the first and second moment estimates, respectively. In addition to exploring the number and sizes of neutron layers, the drop-out fraction, and learning rate parameters, other so-called hyper-parameters explored such as the batch size and the epoch number. No hyper-parameter showed significant impact on the quality of the training and outputs. The largest sensitivity to the overall quality and performance of the NN model came from the type, size, and scope (in terms of the manual gating) of the training-data used for the supervised training.

\subsubsection{Manual data reduction and defining the training data}

\begin{figure*}[htb]
    \centering
    \includegraphics[width=\textwidth]{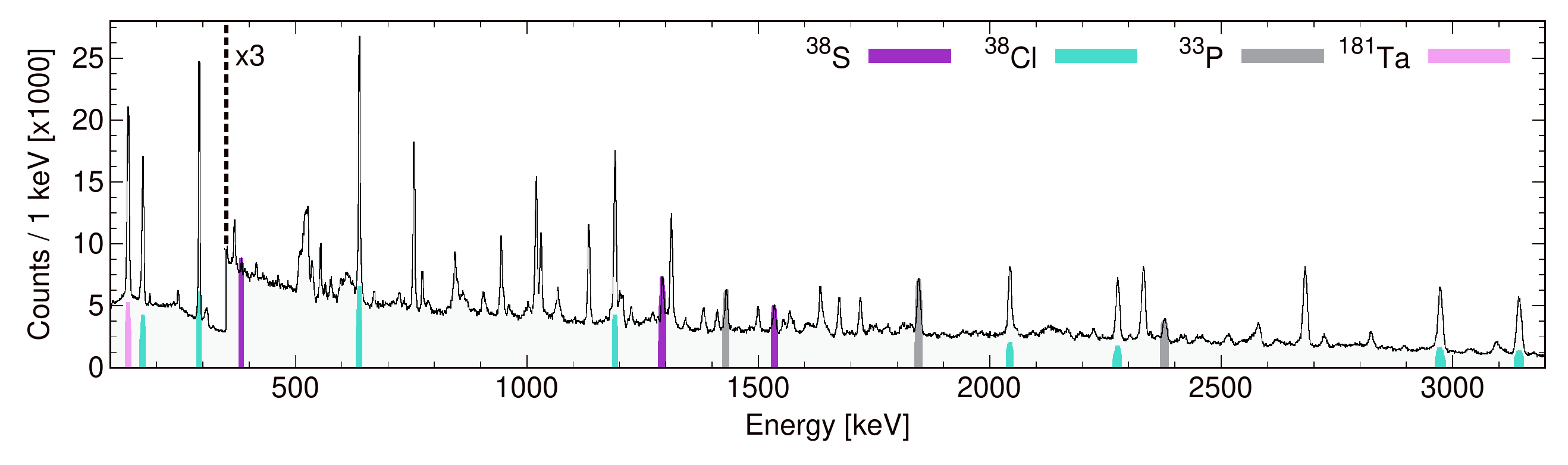}
    \caption{The total spectrum of $\gamma$ rays encompassed within the manual gating regions (see text and Fig.~\ref{fig:fig1}). The labeled events used for the NN model training data set are distinguished by the colored regions. Note that for the labeling of the $^{38}$Cl and $^{181}$Ta $\gamma$-ray transitions, a down-sampling of x4 was used, hence, their reduced size relative to the total counts in spectrum.}
    \label{fig:fig2}
\end{figure*}

The NN model was able to be trained, under supervision, by labeling a sub-set of data based on $\gamma$-ray energy. This was possible because there are known transitions for both the recoil of interest, $^{38}$S~\cite{ref:For94,ref:Oll04,ref:Wang2010,ref:Lun16}, as well as for $^{38}$Cl~\cite{ref:Lub19}, $^{33}$P~\cite{ref:Lub18}, and $^{181}$Ta~\cite{ref:ensdf}. Fig.~\ref{fig:fig2} and Table~\ref{tab:tab1} show and list, respectively, the $\gamma$-ray energies used for labeling the NN training data. Input values that were associated with a $^{38}$S $\gamma$-ray energy $\pm1 - 2$~keV were labeled with a value $\equiv$1. Other types of data with the requisite $\gamma$-ray energies $\pm1 - 2$~keV were similarly labeled as $\equiv$0. Note that: i) not all $\gamma$-rays with the appropriate energies were labeled and used for training (additional details below), and ii) $\gamma$-ray spectra inherently contain backgrounds from various sources as shown in Fig.~\ref{fig:fig2}. Due to the latter, generation of a ``clean'' data set for training was not possible. As a result, there were large contributions to mislabeled events, ranging anywhere from $\sim$10\% to $>50$\% for each of the $\gamma$-ray energies included in the training data set. 

\begin{table}[htb]
\caption{\label{tab:tab1}%
Energies and numbers of known $\gamma$-ray transitions that were included in the manual gating, selected for labeling, and used as NN model training data. }
\begin{ruledtabular}
\begin{tabular}{ccccc}
 & $E_{\gamma}$ (keV) & \# labeled & label & Refs. \\
\colrule
\multirow{3}{*}{$^{38}$S} & 384 & 8600 & \multirow{3}{*}{$\equiv1$} & \multirow{3}{*}{~\cite{ref:For94,ref:Oll04,ref:Wang2010,ref:Lun16}}\\
 & 1293 & 18460 & &\\
 & 1535 & 11940 & &\\
 & & & &\\
\multirow{8}{*}{$^{38}$Cl} & 171 & 20610 & \multirow{8}{*}{$\equiv0$} & \multirow{8}{*}{~\cite{ref:Lub19}}\\
 & 292 & 21110 & &\\
 & 638 & 10210 & &\\
 & 1190 & 7260 & &\\
 & 2044 & 4790 & &\\
 & 2275 & 5880 & &\\
 & 2972 & 6530 & &\\
 & 3142 & 5630 & &\\
  & & & &\\
\multirow{3}{*}{$^{33}$P} & 1432 & 14840 & \multirow{3}{*}{$\equiv0$} & \multirow{3}{*}{~\cite{ref:Lub18}} \\
 & 1848 & 21250 & &\\
 & 2379 & 12900 & &\\
  & & & &\\
\multirow{1}{*}{$^{181}$Ta} & 136 & 25910 & $\equiv0$ & ~\cite{ref:ensdf}\\
\end{tabular}
\end{ruledtabular}
\end{table}

A manual reduction of the data was used to pre-define a region of events for inclusion into the NN model for training data selection as well as for the analysis which followed. The manual selections aided in further reducing random coincidences, background events, and hence, mislabeled training data, over a timing-coincidence requirement alone. Also, the manual gates were aimed at removing an initial overwhelming number of events from $^{38}$Ar and $^{38}$Cl that were not interfering with the cleanliness of the $^{38}$S spectra. Therefore, in addition to a requirement that events have a time relation of $|\Delta T_{\gamma -PGAC}|<200$~ns, selections were also derived from the $E_{1}$-$E_{2}$, $E_{1}$-$E_{3}$, and $E_{1}$-$x$ 2-dimensional spectra as shown in Fig.~\ref{fig:fig1}. The energy selections represented by the grey data points in Figs.~\ref{fig:fig1}(b) and (d) were able to remove all events linked to the $^{38}$Ar isobar, as well as a large portion of the $^{38}$Cl recoil events. The gates also reduced the number of random $\gamma$-ray events caused by scattered $^{22}$Ne primary beam. The selection region shown in Fig.~\ref{fig:fig1}(f) on $E_{1}$-$x$ removed most of the mass ambiguities. However, the $A/q\approx 33/7$ region was purposefully not removed so as to ensure enough $^{33}$P events remained for labeling and inclusion in the NN model training. In total, $\sim5.5\times10^{6}$ $\gamma$ rays remained after the manual selections (Fig.~\ref{fig:fig2}). Included within those data were the $\sim$195k events labeled for training, $\sim$39k $^{38}$S, $\sim$82k $^{38}$Cl, $\sim$49k $^{33}$P, and $\sim$26k Ta (Table~\ref{tab:tab1} and ~\ref{fig:fig2}). In order to provide better balance to the training data, the labeling of the $^{38}$Cl and $^{181}$Ta data was down-sampled by a factor of 4 each.

\begin{figure*}[htb]
    \centering
    \includegraphics[width=0.48\textwidth]{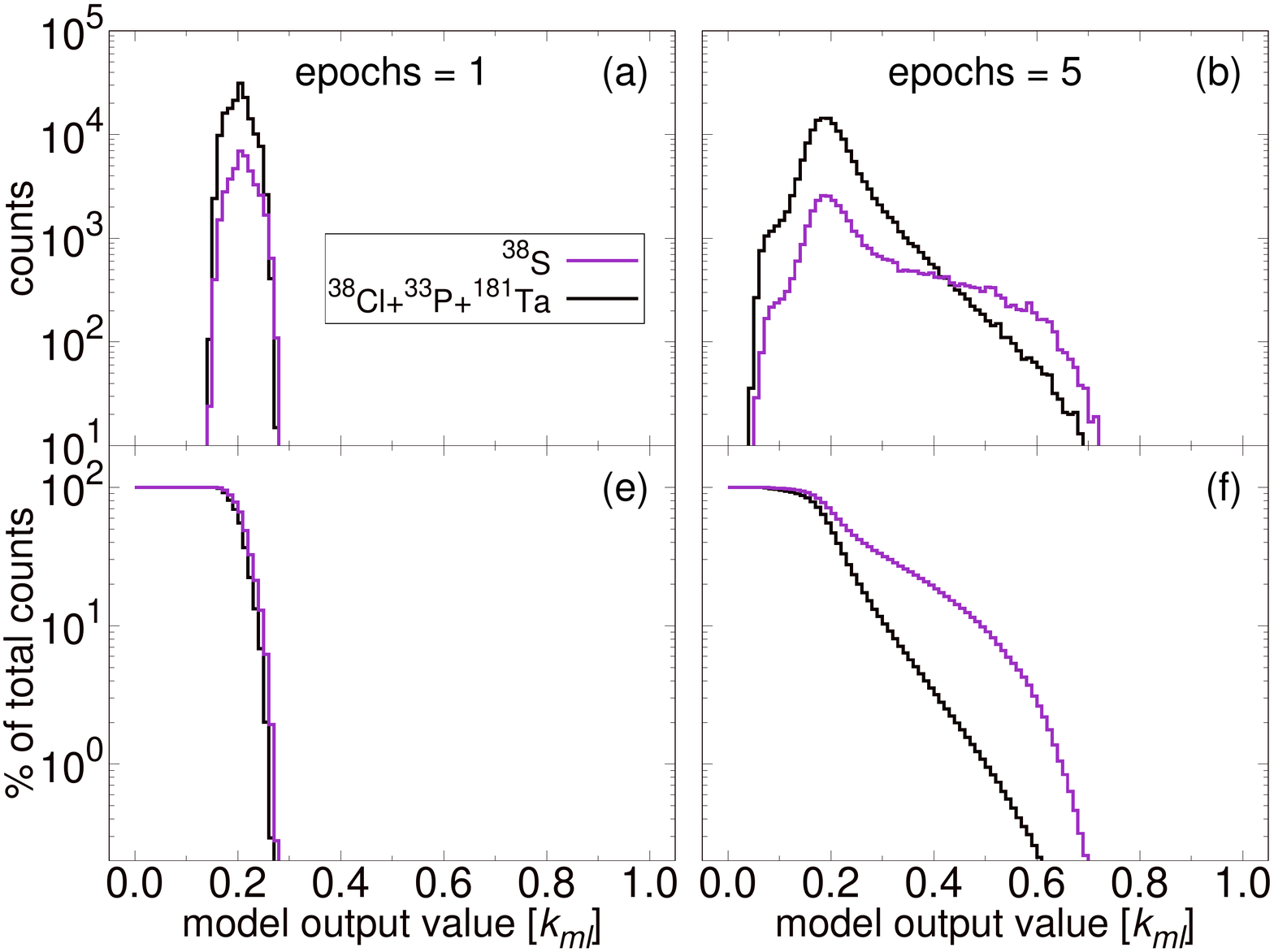}
    \hspace{-0.95cm}
    \includegraphics[width=0.48\textwidth]{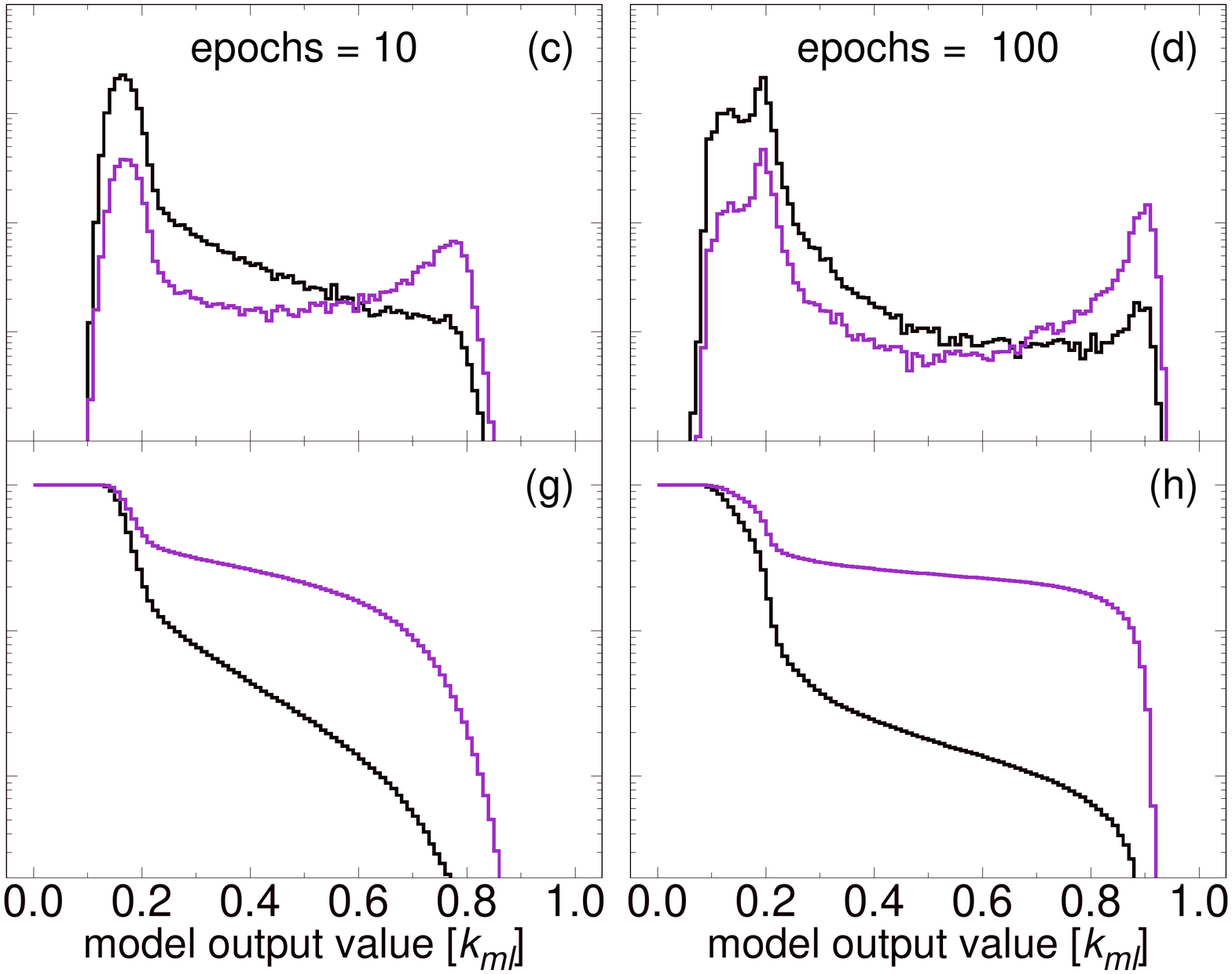}
    \caption{(a - d) The evolution of the distribution of model output values, $k_{ml}$, for the training data at snapshots throughout the NN model training at completed epoch cycles (1, 5, 10, and 100). Purple lines correspond to the $k_{ml}$ values of the labeled $^{38}$S training data and the black lines are the corresponding values of the sum of the $^{38}$Cl, $^{33}$P, and $^{181}$Ta labeled training data. (e - h) The fraction of the total number of integrated counts (given in \%) over the integral range starting from the $k_{ml}$ value given on the x-axis through $k_{ml}=1$, i.e. $\int_{k_{ml}}^{1}dN$/$\int_{0}^{1}dN$.}
    \label{fig:fig3}
\end{figure*}

\begin{figure}[htb]
    \centering
    \includegraphics[width=0.48\textwidth]{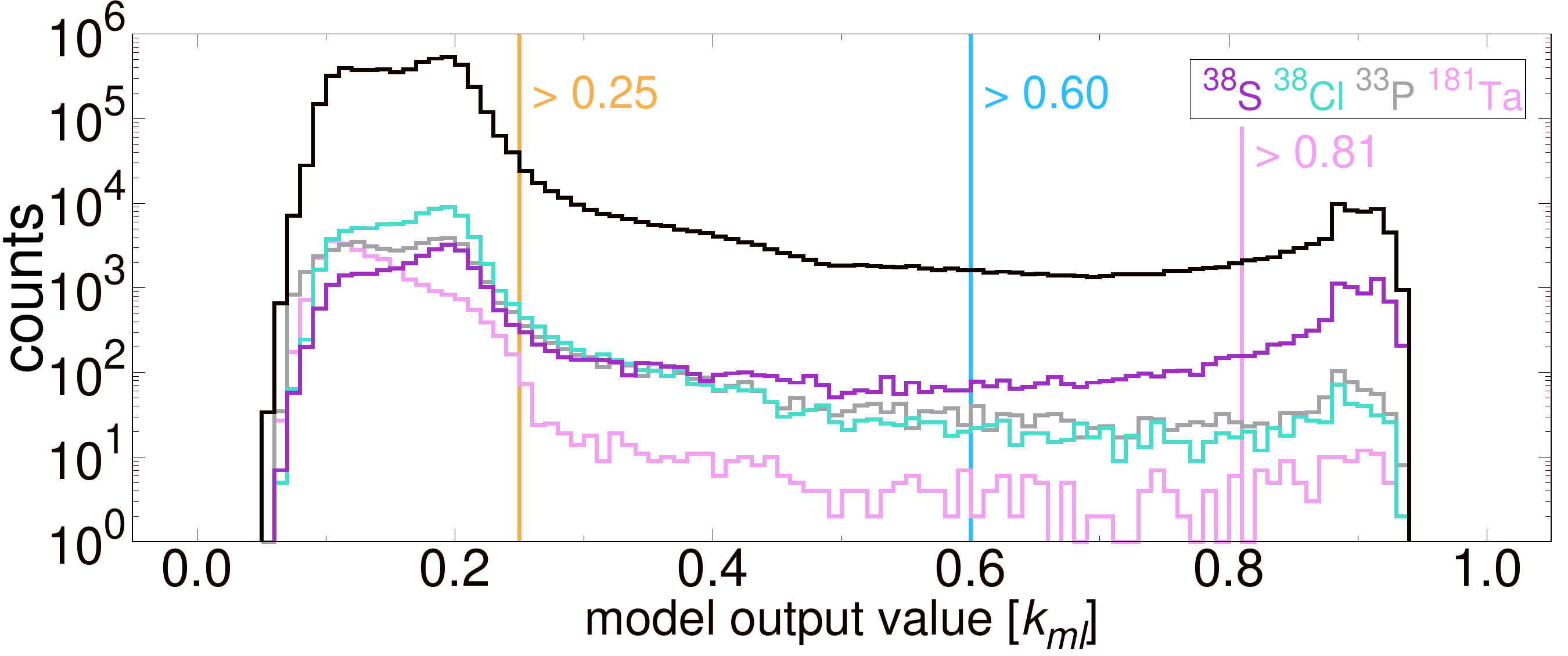}
    \caption{The distribution of the model output values, $k_{ml}$, for the fully-trained NN model (epochs = 200). The black line represents the full complement of data encompassed within the manual gating regions, both labeled and un-labeled. The colored lines correspond to the final distributions of each of the separately labeled training data only (purple - $^{38}$S, blue - $^{38}$Cl, grey - $^{33}$P, pink - $^{181}$Ta). The yellow ($>0.25$), blue ($>0.60$), and pink ($>0.81$) vertical lines identify the lower-limit $k_{ml}$ integral values, where the upper limit was defined $=1$, applied throughout the $\gamma$-ray analysis (see text for additional details).}
    \label{fig:fig4}
\end{figure}

\subsubsection{NN model training \& output}
The training of the NN model specifically refers to the adjustment and eventual optimization of the 1322 model parameters described above. The supervised training of the NN model, whereby we utilized our labeled data, was completed by cycling through the labeled training data 200 times (200 epochs). The loss calculations and backward propagation were calculated and carried out in batches of 500 pieces of data. Changes in the epoch and batch size parameters were explored but did not significantly alter the model performance. Throughout the training procedure, the model output value $k_{ml}$ was rounded to its nearest integer (0 or 1) prior to the loss calculation. Various iterations into the type and scope of the manual data selection, as described above, were explored in order to reach the adopted parameters for the NN model. Metrics typically used in defining the goodness of trained machine-learning models were not crucial in the determination of the applicability of the model in this case. This is in part due to the known mislabeling of a large fraction of training events which would lead to ambiguities in most standard metrics. 

To first order, the distribution and trends of the $k_{ml}$ values provided straight-forward visual feedback of the NN model's ability to separate $^{38}$S data from the rest. This can be visualized through the evolution of the $k_{ml}$ values for the training data throughout the actual training process (Fig.~\ref{fig:fig3}). In Figs.~\ref{fig:fig3}(a - d) the distributions for $k_{ml}$ are shown at the conclusion of different epoch numbers (1, 5, 10, and 100) for the labeled $^{38}$S training data (purple lines) as well as the sum of all other labeled training data (black lines). As expected, after a single pass through the data (epochs = 1) both types of labeled data have similar distributions. As the epoch number grows, the $^{38}$S data starts trending towards $\equiv1$ faster than the other data types, and even after only 10 epochs, a peak forms for the $^{38}$S data above $k_{ml}>0.5$. By the completion of 100 epochs, the distributions are different and a sub-set of data, peaking towards $k_{ml}=1$ identifies the $^{38}$S data within the model. Fig.~\ref{fig:fig4} also shows the breakdown of the individually labeled components for the training data $k_{ml}$ distributions with the fully-trained model (epochs = 200). One sees that little evolves between epochs 100 - 200. Also, the behavior of each non-$^{38}$S type is similar.

Beyond the distribution of the $k_{ml}$ values, the integrated number of counts for a specific $\gamma$-ray energy between a starting value of $k_{ml}=X$ up to $k_{ml}=1$, or $k_{ml}>X$, was also calculated throughout each training. The behavior of the fractional counts remaining for the $^{38}$S $\gamma$ rays, as well as the combined $\gamma$ rays of the contaminants, are given relative to the total (in percent) in Figs.~\ref{fig:fig3}(e - h). The trends mimic those of the $k_{ml}$ distributions as expected and highlight the unique persistence of the $^{38}$S counts for increasing the lower limit on the integration.

The entirety of the data that was encompassed within the manual-cut regions (Figs.~\ref{fig:fig1} and~\ref{fig:fig2}), not just the data that was labeled, was similarly evaluated by the fully-trained NN model. The distribution of all of the output $k_{ml}$ values is shown by the black line in Fig.~\ref{fig:fig4}. The complete data set and the labeled training data were used to define the $k_{ml}$ values applied in the final analysis. Contaminant $\gamma$-ray yields showed the most dramatic reduction in counts when $k_{ml}\gtrsim0.25$, i.e. the inclusion of counts between $0.25\leq k_{ml}\leq1$. The $^{38}$S yields varied only slightly through this transition point. In addition, the $^{38}$S recoils remained nearly constant over $0.25\lesssim k_{ml} \lesssim 0.8$, while contaminant yields continued to reduce. Above $k_{ml}\gtrsim0.8$ all events showed diminishing yields.

Finally, the validity of each trained model was also evaluated by the quality of the $\gamma$-ray singles spectra that they could generate. Specifically, the labeled training data was used to check ratios between the summed areas of known transitions in $^{38}$S~\cite{ref:For94,ref:Oll04,ref:Wang2010,ref:Lun16} compared to the relative summed areas of the known transitions in $^{38}$Cl~\cite{ref:Lub19} and $^{33}$P~\cite{ref:Lub18}. For consistency, the comparison between different trained models was made at a $k_{ml}$ value in which the area of the $^{38}$S 1293-keV transition was $\equiv5000$ counts. The $\gamma$-ray spectra from each NN model training were also visually inspected and overlaid with previously generated spectra in order to qualify background suppression. The resulting spectra are presented in Section~\ref{sec:res}.

\subsection{Additional prompt $\gamma$-ray information}
In addition to a $\gamma$-ray singles spectrum for a specific $k_{ml}$ value, a recoil-$\gamma$-$\gamma$ coincidence matrix was also generated from events which had $\gamma$-ray multiplicities $M_{\gamma}>1$. To be included in the matrix, a pair of $\gamma$ rays must have had a relative timing relation of $<200$~ns and both have been within the accepted $k_{ml}$ cut range. The cut range $k_{ml}>0.25$ was primarily used in the analysis due to its compromise between cleanliness and statistics. In cases where contaminants may have been present, including those from $^{38}$Cl in particular, coincidences were checked with data from the more stringent $k_{ml}>0.60$ cut. Statistics limited the observation of any higher-order $\gamma$-ray multiplicity studies.

The semi-alignment of magnetic sub-states provided by the near-symmetric fusion evaporation reaction was leveraged to inform on some transition multipolarities. The ratio of the $\gamma$-ray singles yields, $R_{\theta_2/\theta_1}$, was extracted from data included within the $k_{ml}>0.60$ cut. Although GRETINA has essentially a continuous angular coverage, only two angular bins were used due to the limited statistics with one centered around $\theta_2 = 149^{\circ}$ and the other centered around $\theta_1 = 90^{\circ}$. An energy-dependent efficiency curve was determined independently for each of the corresponding angular binning regions from the $\gamma$-ray source data. Above $E_{\gamma}>500$~keV the ratio of the two efficiency curves to one-another was uniform. A systematic uncertainty of 5\% was adopted for that energy region. Below this energy, a systematic uncertainty of 10\% was adopted due to an increase in the sensitivity of the efficiency-curve fit parameters to the $\theta_1 = 90^{\circ}$ source data. The final uncertainty on $R_{149^{\circ}/90^{\circ}}$ also included those from statistics. Stretched-quadrupole ($\Delta J=2$, $L=2$) transitions are expected to reside above $R_{149^{\circ}/90^{\circ}}\gtrsim1.2$, while the dipole $\Delta J\leq1$, $L=1$ counterpart transitions are more probable for $R_{149^{\circ}/90^{\circ}}\lesssim 1$ values. The known 1293-keV and 1535-keV $^{38}$S $\gamma$-rays, both having $\Delta J=2$, gave $R_{149^{\circ}/90^{\circ}}~\gtrsim$~1.2, in agreement with expectations.

\section{\label{sec:res}Experimental Results}

\subsection{Selection of $^{38}$S $\gamma$-ray transitions}
\begin{figure*}[htp]
    \centering
    \includegraphics[width=\textwidth]{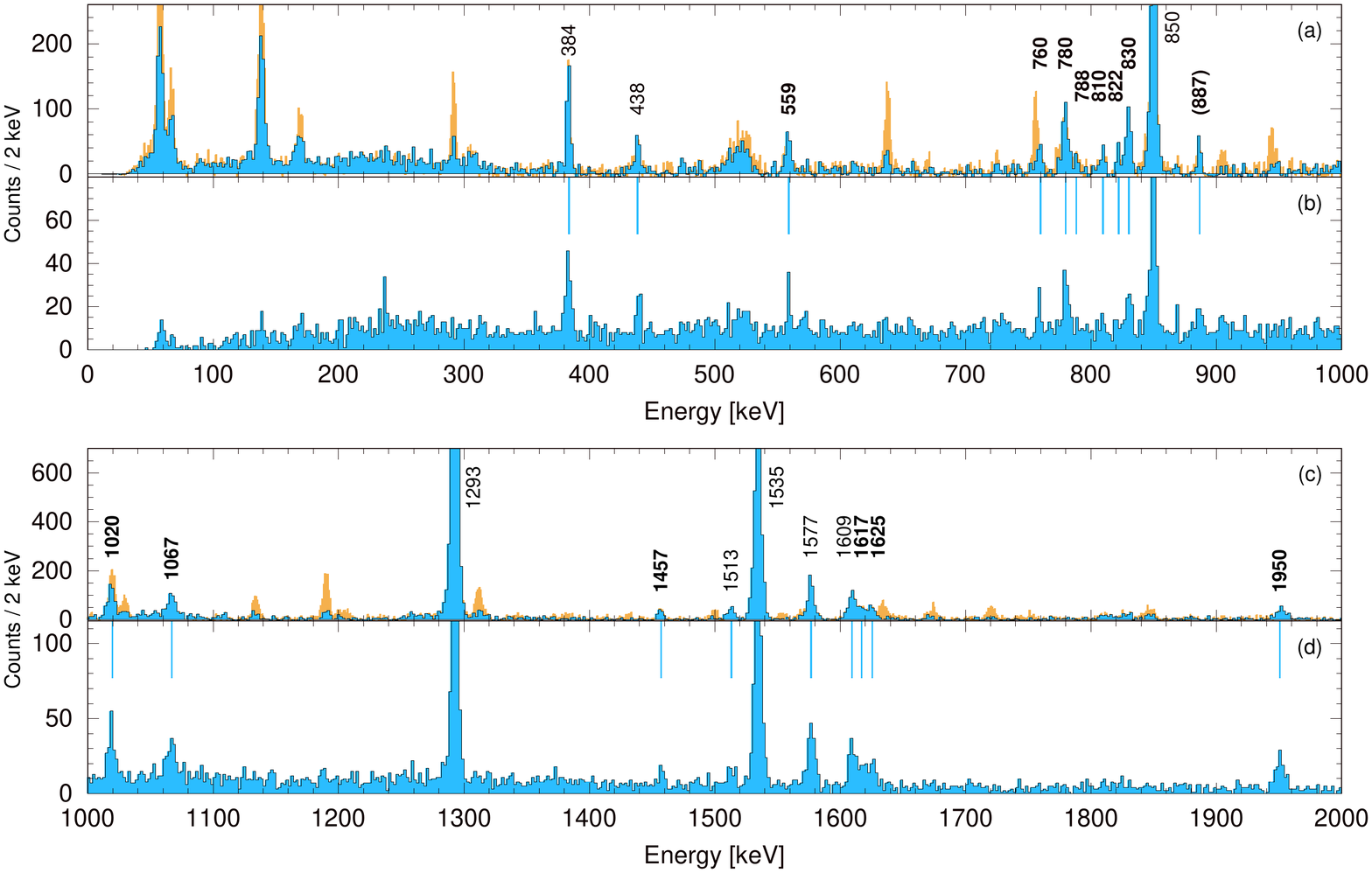}\\
    \vspace{-0.8in}
    \includegraphics[width=\textwidth,trim={0 4in 0 0},clip]{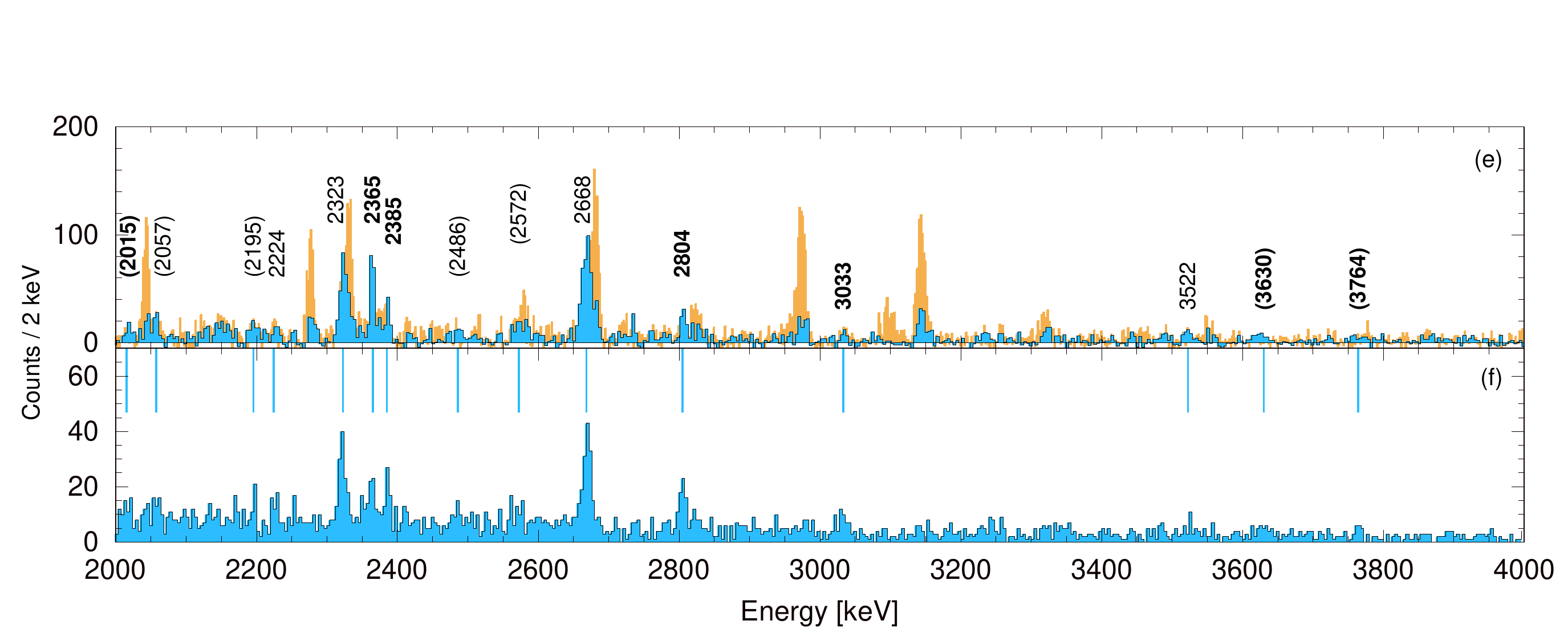}
    \vspace{0in}
    \caption{(a),(c),(e) Background-subtracted $\gamma$-ray singles spectra from a model output cut of $k_{ml}>0.6$ (blue) overlaid on a cut of  $k_{ml}>0.25$ (yellow). All transitions belonging to $^{38}$S are labeled by their energies while those in bold have been newly determined in the present work. Parenthesis indicate transitions that were unable to be placed in the level scheme (Fig.~\ref{fig:fig6}) and may be considered tentative. The recoil-$\gamma$-$\gamma$ spectrum ($k_{ml}>0.6$ cut) for each $^{38}$S transition labeled in (a), (c), and (e), were combined to generate the inclusive summed spectra shown in (b), (d), and (f). The blue vertical lines correspond to the labeled energies of the above histogram.}
    \label{fig:fig5}
\end{figure*}

 The $\gamma$-ray transitions attributed to $^{38}$S based on the present analysis are labeled by their observed energies in the $\gamma$-ray singles spectra of Fig.~\ref{fig:fig5} and they are also listed in Table~\ref{tab:tab2}. The $\gamma$-ray energies carry an uncertainty of $\sim$0.5 - 1~keV. Transitions in $^{38}$S were identifiable by showing little change in their yields between the spectra with the $k_{ml}>0.25$ cut versus the $k_{ml}>0.60$ cut as shown in Figs.~\ref{fig:fig5}(a), (c), and (e). Spectra generated from the $k_{ml}>0.60$ cut (blue) were far-removed of the contaminant lines, in particular those from $^{38}$Cl, which appear prominently in the $k_{ml}>0.25$ selection (yellow) but diminish largely for the $k_{ml}>0.6$ selection. Note that a smooth background has been subtracted from the singles spectra in Figs.~\ref{fig:fig5}(a), (c), and (e). The summed spectra, including each individual recoil-$\gamma$-$\gamma$ spectra ($k_{ml}>0.6$), shown in the lower half of each panel of Fig.~\ref{fig:fig5}, provide additional support of the associated $^{38}$S transitions. The relative intensities of the $^{38}$S transitions were determined from the singles spectra of the $k_{ml}>0.60$ data. They are given in Table~\ref{tab:tab2}, normalized to the intensity of the 1293-keV line ($\equiv1000$). Uncertainties on the relative intensities were dominated by statistics over the systematic contributions from the peak fitting and background subtraction ($<5$\%) or the energy-dependent efficiency correction ($<5$\%).
 
\begin{figure*}[htp]
    \centering
    \includegraphics[width=\textwidth]{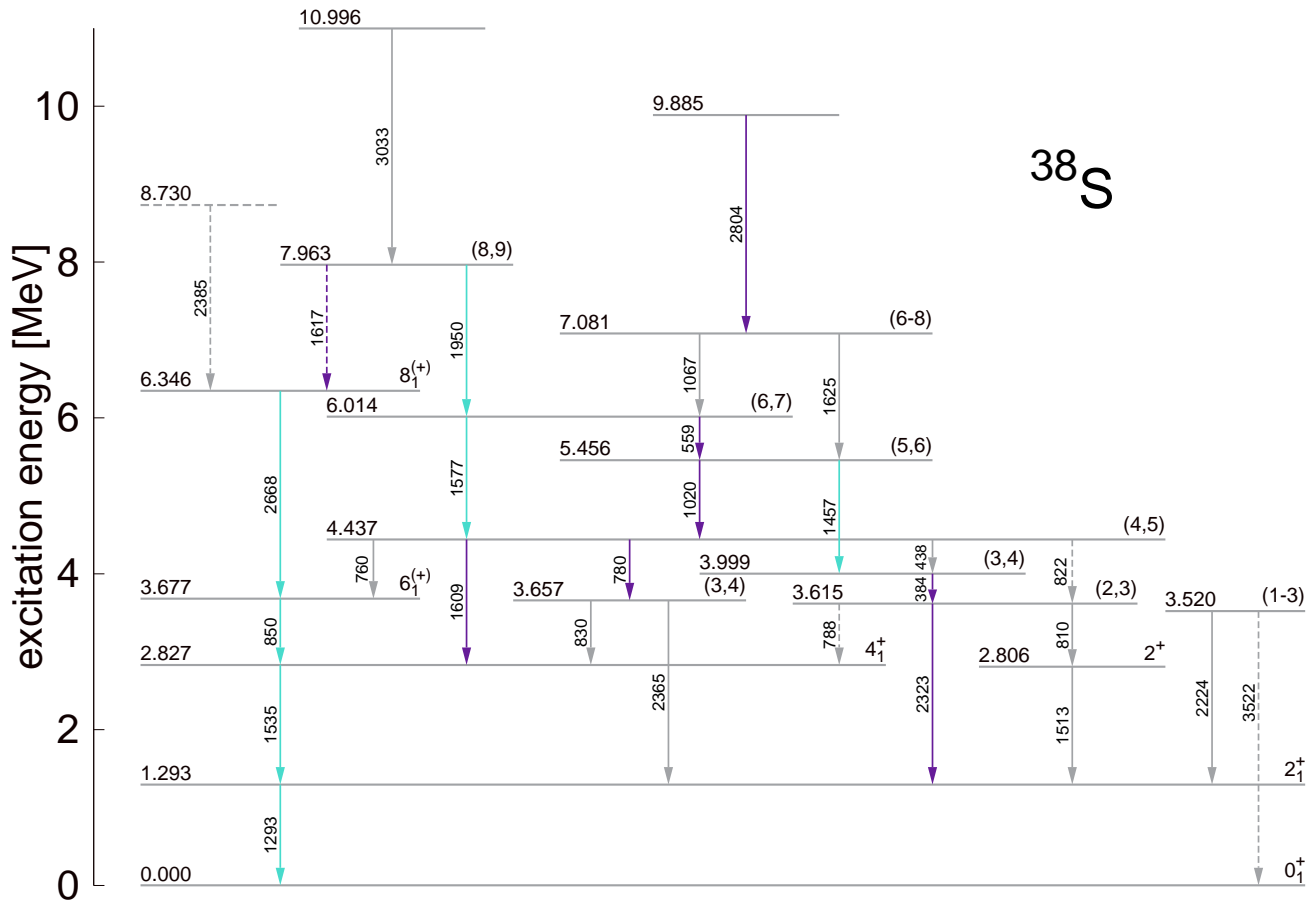}
    \caption{The level scheme of $^{38}$S based upon the present work and other in-beam measurements. Dashed levels and transitions indicate tentative placement. Transitions in blue ($L=2$) and purple ($L=1$) reflect multipolarities based on their $R_{149^{\circ}/90^{\circ}}$ values (Fig.~\ref{fig:fig7} and Table~\ref{tab:tab2}).}
    \label{fig:fig6}
\end{figure*}

\begin{figure}[htp]
    \centering
    \includegraphics[width=0.5\textwidth]{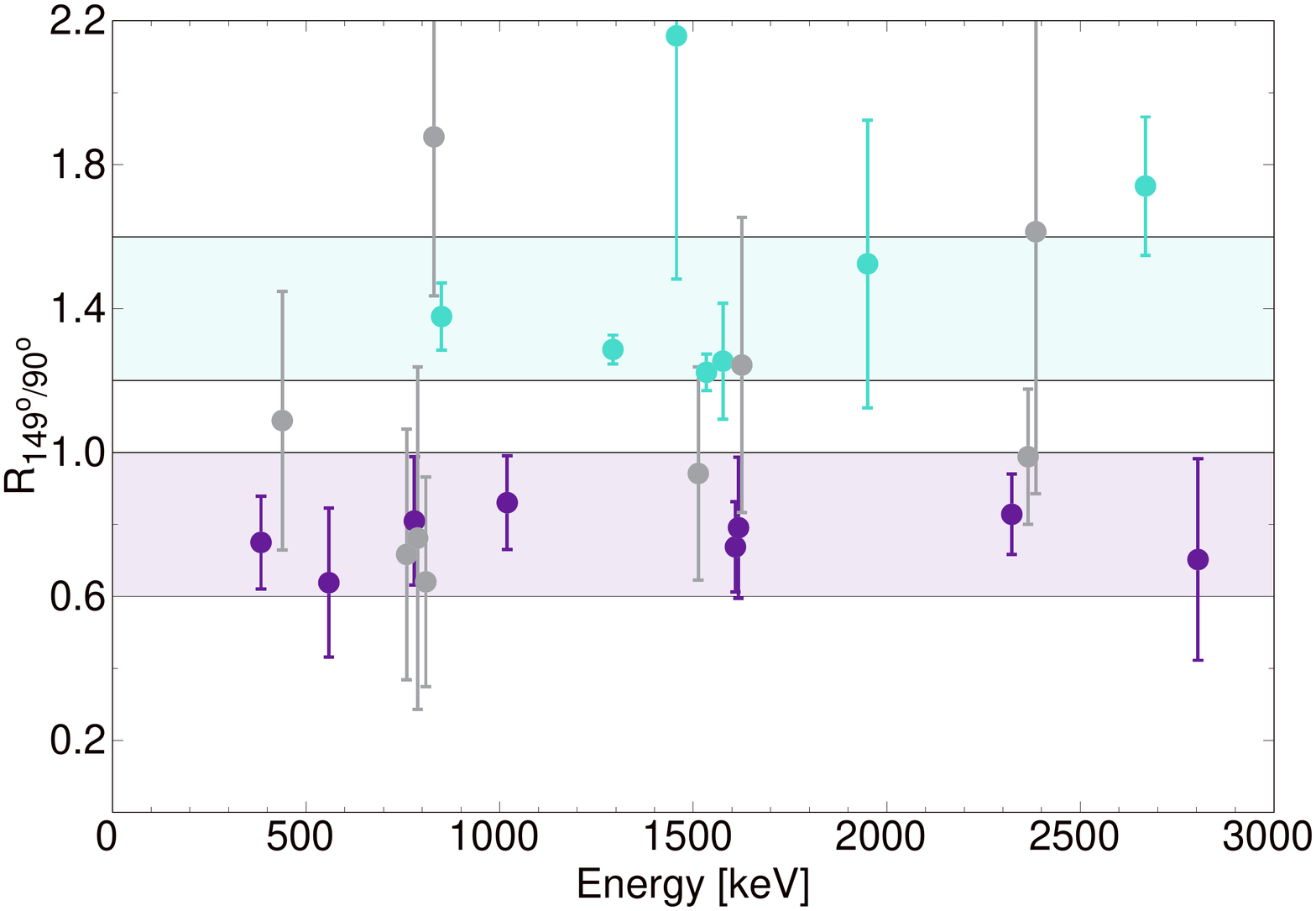}
    \vspace{-0in}
    \caption{The ratio of the extracted $\gamma$-ray yields centered around $\theta_2=149^{\circ}$ by that around $\theta_1=90^{\circ}$, $R_{149^{\circ}/90^{\circ}}$, for identified transitions in $^{38}$S as a function of their energy. The colored data points correspond to either fully-stretched quadrupole transitions residing above $\approx1.2$ (blue for multipolarity $L=2$) or dipole transitions residing below $\approx1$ (purple for multipolarity $L=1$). Grey data points were not assigned multipolarities primarily due to large uncertainties.}
    \label{fig:fig7}
\end{figure}

\subsection{$^{38}$S level scheme}
An updated $^{38}$S level scheme incorporating the present data is shown in Fig.~\ref{fig:fig6}. Some details pertaining to the construction of specific levels or locations of transitions are given below. In general, the placement of transitions into the $^{38}$S level scheme was done through the use of $\gamma$-ray and excited level energy summations, the relative $\gamma$-ray intensities (Table~\ref{tab:tab2}), and where possible, utilizing the recoil-$\gamma$-$\gamma$ coincidence data (Fig.~\ref{fig:fig8}). Dashed transitions or levels in Fig.~\ref{fig:fig6} identify cases where the coincidence data was inconclusive or where only the energy-sums were used. Newly assigned or speculative $J^{\pi}$ information was extracted from the $R_{149^{\circ}/90^{\circ}}$ values (Fig.~\ref{fig:fig7} and Table~\ref{tab:tab2}), as well as transition selection rules and the propensity of the fusion evaporation reaction mechanism to populate higher-$J$ or yrast states with increasing excitation energy.

\subsubsection{\label{subsec:yrast}The yrast even-$J$ levels}
\begin{figure}[htp]
    \centering
    \includegraphics[width=0.48\textwidth]{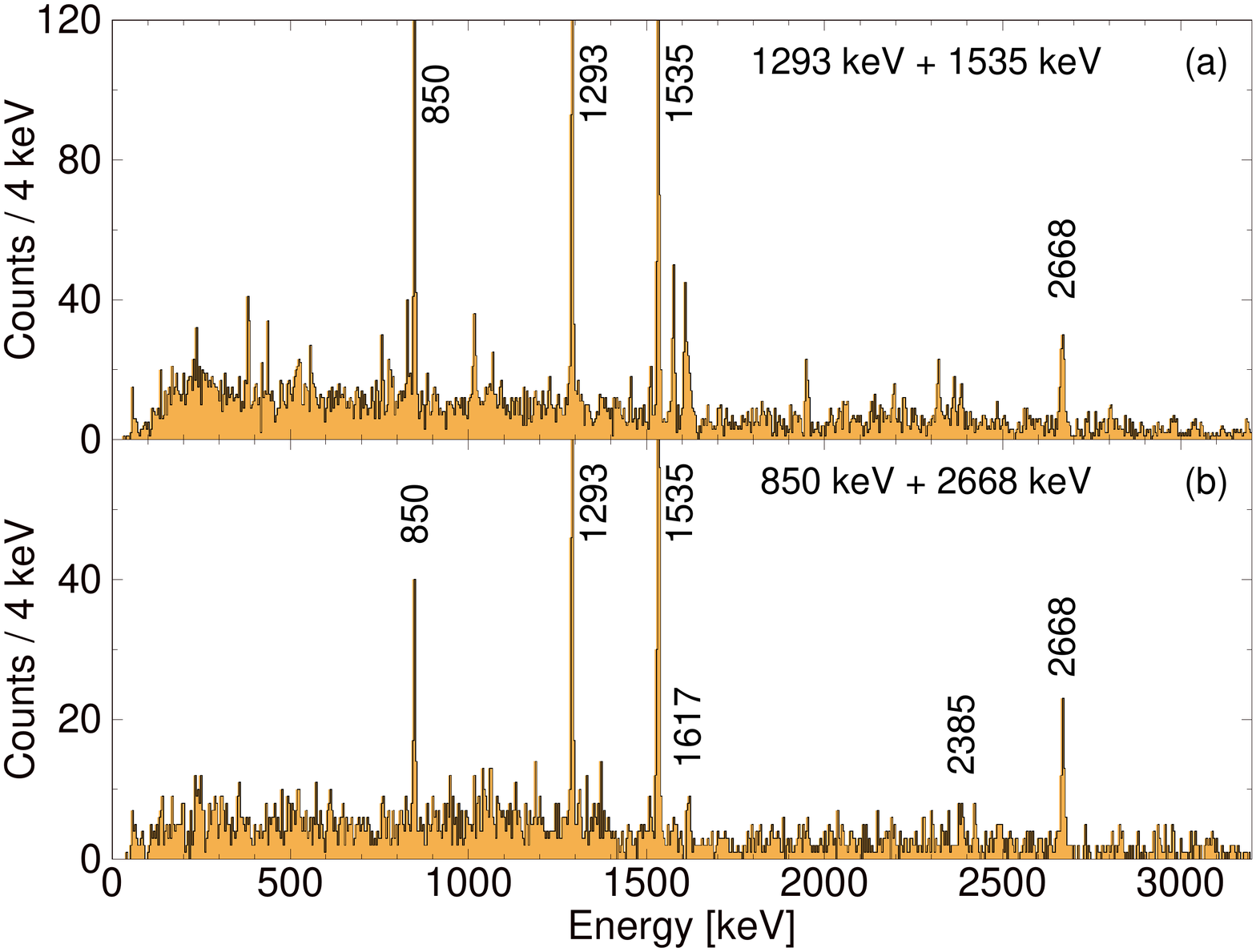}
    \includegraphics[width=0.48\textwidth]{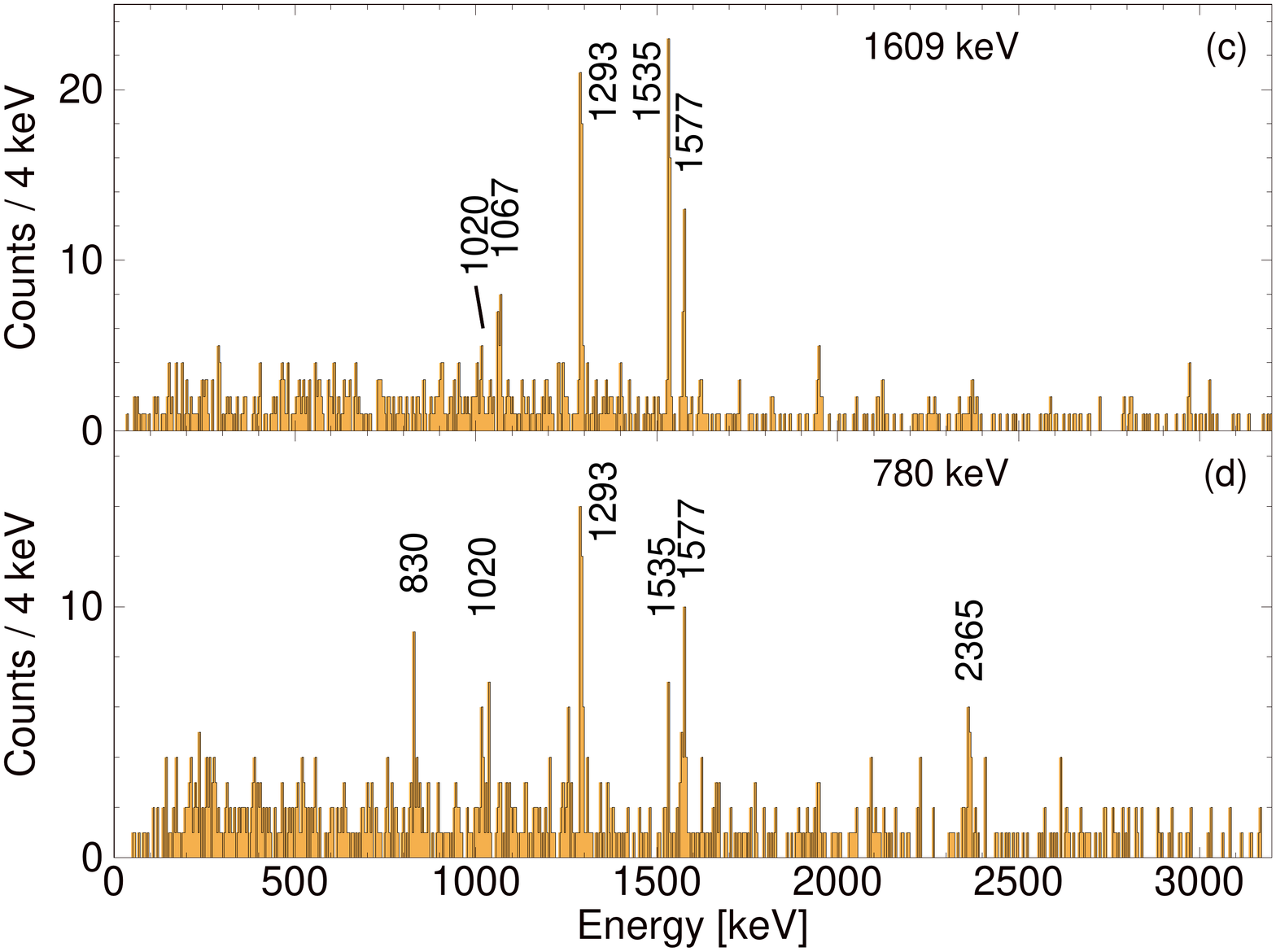}\\
    \includegraphics[width=0.48\textwidth]{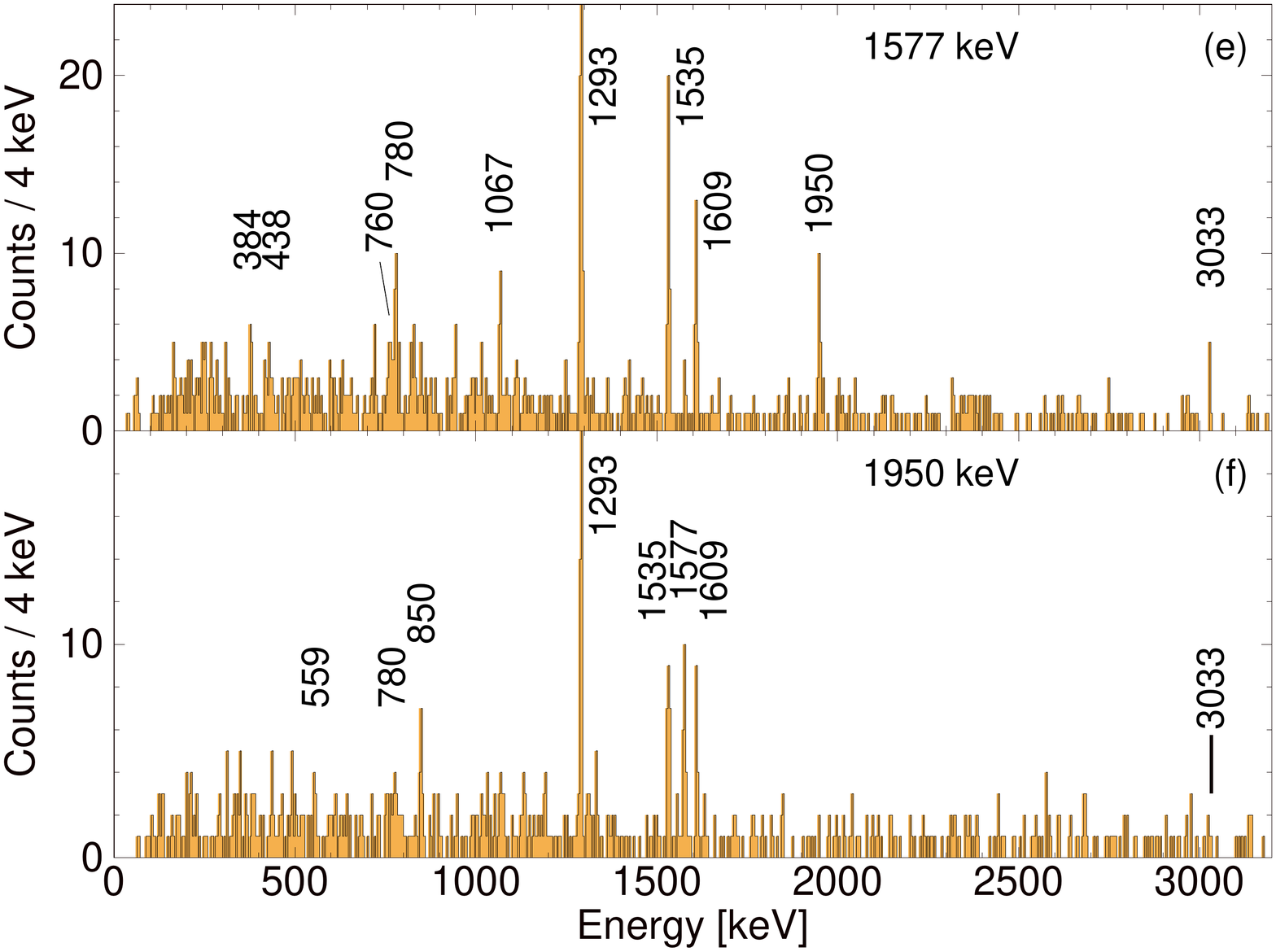}\\
    \caption{Projected recoil-$\gamma$-$\gamma$ coincidence spectra for $k_{ml}>0.25$. The labels provide the $\gamma$-ray energy or energies selected for the projection. The "$+$" sign represents the summation of two separate recoil-$\gamma$-$\gamma$ spectra. Only some of the transitions of interest have been labeled in each spectra. No background subtractions have been applied.}
    \label{fig:fig8}
\end{figure}

The energies and $J^{\pi}$ values for the even-$J$ yrast levels up to $J^{\pi}=4^+$ have been firmly established (0.000~MeV -- 0$^+$, 1.293~MeV -- 2$^+$, and 2.827 MeV -- 4$^+$). A tentative $J^{\pi}=(6^{+}_{1})$ assignment to the 3.677-MeV level was made previously based on speculative ($t$,$p$) angular distributions~\cite{ref:Dav85}, as well as observation of sizeable population and a predominant decay branch to the 4$^+_1$ level by various in-beam reaction work~\cite{ref:For94,ref:Oll04,ref:Wang2010,ref:Wang,ref:Lun16,ref:Grocutt2022}. The $\gamma$-ray coincidence data and the relative intensities extracted in the present work agree with the established 1293-, 1535-, and 850-keV $\gamma$-ray cascade and arrangement [Figs.~\ref{fig:fig8}(a) and (b)].  $R_{149^{\circ}/90^{\circ}}\gtrsim1.2$ values were extracted for each transition, including the 850-keV $\gamma$ ray, showing consistent stretched quadrupole multipolarity throughout the cascade (Fig.~\ref{fig:fig7} and Table~\ref{tab:tab2}). The assignment of the 3.677-MeV level has therefore been modified to $J=6^{(+)}_1$, where positive parity is most likely.

A coincidence relationship was observed between the known yrast $J^{\pi}=6^{(+)} \rightarrow 4^+ \rightarrow 2^+ \rightarrow 0^+$ sequence and the 2668-keV $\gamma$ ray [Figs.~\ref{fig:fig8}(a) and (b)]. First identified in Refs.~\cite{ref:Wang2010,ref:Wang}, the 2668-keV transition has now been placed to directly feed the 3.677-MeV level from a new level at 6.346 MeV. No additional transitions were observed to decay from this new level suggesting a decay branch of near $\sim$100\% to the 6$^{(+)}_1$ level. The efficiency corrected intensities for the 1535- and 850-keV transitions relative to the 1293-keV [$\equiv1.00(12)$] were 0.98(12), 0.85(11) in the 2668-keV gated $\gamma$-$\gamma$ coincidence spectrum. The extracted $R_{149^{\circ}/90^{\circ}}=1.74(19)$ value of the 2668-keV $\gamma$ ray shows a clear quadrupole multipolarity. Hence, the new level at 6.346 MeV has been assigned as the yrast $J=8^{(+)}_1$ level. Similar to the 3.677-MeV level, this level is likely $\pi=+$.

Extending the even-$J$ yrast sequence beyond $J=8$ with any certainty proved difficult. Exploration of the summed and individual coincidence spectra for the yrast even-$J$ transitions revealed two possible candidate $\gamma$ rays at 1617-keV and 2385-keV [Fig.~\ref{fig:fig8}(b)]. Any coincidence relationship between these two lines was uncertain due to the limited statistics. Coincidence with the 2668-keV transition in both cases led to their tentative placements directly feeding into the $J^{\pi}=8^{(+)}$ 6.346-MeV level. The tentative level at 8.730 MeV has no spin-parity suggestion as the multipolarity information on the 2385-keV $\gamma$-ray is inconclusive [$R_{149^{\circ}/90^{\circ}}=1.61(73)$] though $J\geq8$ is most likely. The placement of the 1617-keV transition generated a level with an energy sum of $<$2 keV within that of the 7.963-MeV level established by the 1950-keV transition (see sub-section~\ref{subsec:1950} below). The 1617-keV $\gamma$-ray's $R_{149^{\circ}/90^{\circ}}<1.0$ value favored a dipole multipolarity and the 7.963-MeV level having $J=(8,9)$ is consistent with the 1950-keV cascade.

\subsubsection{The 2.806-MeV, 3.520-MeV, 3.615-MeV and 3.999-MeV levels}
The energies of the 2.806-, 3.520-, 3.615-, and 3.99-MeV levels were previously established with $J^{\pi}$ values of $2^+_2$, ($1-3$), ($2^+,3^+$) and ($3^+,4^+$), respectively (see Fig.~6.46 of Ref.~\cite{ref:Wang} for example). The doublet of states around 2.8 MeV was cleared up definitively in Ref.~\cite{ref:War87} and a recent work confirmed the assignment of the 2.806-MeV level as $J^{\pi}=2^+_2$~\cite{ref:Longfellow2021}. The 3.520-MeV level was first identified in the $\beta$-decay of $^{38}$P~\cite{ref:Duf86} leading to a limit on $J$ from the ($2^-$) ground state. The same level was also weakly populated in the deep-inelastic work of Refs.~\cite{ref:Wang2010,ref:Wang}. While the 2224-keV transition matches the energy of the previous work within errors, it is unclear if the observed weakly-populated 3.522-keV transition is the ground state transition as it is outside the expected energy uncertainty. The 3.615-MeV and 3.999-MeV levels were suggested to be part of a multiplet of levels with $J^{\pi} = 2^+ - 5^+$ (including the 4.437-MeV level discussed below) based on the $(\nu0f_{7/2}\nu1p_{3/2})$ configuration~\cite{ref:Oll04,ref:Wang2010}.

All of the previously observed transitions from these levels have been confirmed in the present work (Fig.~\ref{fig:fig5}). Two new linking transitions of 810-keV and 788-keV were observed to decay from the 3.615-MeV level. The location of 810-keV line was established through the recoil-$\gamma$-$\gamma$ coincidence data with the 1513-keV line. The weak 788-keV transition meets the energy requirements to link the 2.827-MeV and 3.615-MeV levels. It also showed marginal coincidence with the 384-keV line and the lower-lying 1293- and 1535-keV lines. The appearance of the 1535-keV transition within the coincidence spectrum of the 384-keV line also indirectly supported its placement.

No new information on the properties of the $J^{\pi}=2^+_2$ 2.806-MeV level was determined in the present work due to the limited population of this non-yrast state. The $R_{149^{\circ}/90^{\circ}}$ values for the 2323-keV and 384-keV transitions support the sequence of dipole transitions proposed in Refs.~\cite{ref:Oll04,ref:Wang2010,ref:Wang} for the 3.615-MeV and 3.999-MeV levels. There is a propensity for the higher of the two possible $J$-value assignments in each case due to the fusion-evaporation reaction mechanism, however, no direct empirical evidence is available to solidify this point. Therefore, the two levels remain with $J^{\pi}=(2,3)$ and $(3,4)$.

\subsubsection{\label{subsec:3657}The 3.657-MeV level}
A new level was established at an excitation energy of 3.657 MeV and assigned with tentative spin values of $J=(3,4)$. Energy summations and recoil-$\gamma$-$\gamma$ coincidence relations between the 830-keV, 780-keV, and 2365-keV transitions with the known 4$^+_1\rightarrow2^+_1\rightarrow0^+_1$ cascade were used [Fig.~\ref{fig:fig8}(d)]. There were no previously recorded levels having energies consistent with 3.657 MeV, although the 2365-keV was also observed in Refs.~\cite{ref:Wang2010,ref:Wang}. The $\sim$830-keV data observed in Ref.~\cite{ref:Lun16} originated from the 850-keV transition but was shifted in energy due to its $>100$~ps lifetime, hence, it is not the same transition as has been observed here. The tentative spin-parity range came from the limits of the 2365-keV transition to the known $2^+_1$ level and the 830-keV transition to the known 4$^+_1$ level ($2\leq J \leq 4$). The $R_{149^{\circ}/90^{\circ}}=0.81(18)$ value of the 780-keV transition favors $3\leq J \leq 5$ when coupled with the possible $J$ values of the 4.437-MeV level (see subsection~\ref{subsec:4437} below). The $R_{149^{\circ}/90^{\circ}}=1.88(44)$ of the 830-keV line is not obviously consistent with either $J=(3,4)$~$\rightarrow$~$4^+_1$ transition.

\subsubsection{\label{subsec:4437}The 4.437-MeV level}
A key marketplace of the $^{38}$S level scheme has appeared at the 4.437-MeV level. Similar levels were previously identified in the deep inelastic works of Refs.~\cite{ref:Oll04,ref:Wang2010,ref:Wang} at both 4.436-MeV and 4.437-MeV. The former was tentatively assigned $J=(4^+,5^+)$ based on theoretical arguments that the level was a member of the ($\nu0f_{7/2}\nu1p_{3/2}$)$_{J=2-5}$ multiplet. The latter was only postulated based on the placement of an observed 1611-keV transition, though the two levels agreed in energy within uncertainties. A broad peak around 4.43 MeV was also observed in the heavy-ion transfer work of Ref.~\cite{ref:May84} as well as a level at 4.478(22) MeV which was given $J=(3^-,4^+)$ in the ($t,p$) work of Ref.~\cite{ref:Dav85}.

The $4.437 \rightarrow 3.999 \rightarrow 3.615$-MeV decay sequence suggested in Refs.~\cite{ref:Oll04,ref:Wang2010} was confirmed by the recoil-$\gamma$-$\gamma$ coincidence relations between the 384-keV and 438-keV transitions. The coincidence spectra of the 780-keV and 1609-keV lines showed a number of transitions that were also found to be in coincidence with the 438-keV transition, including the 1577-keV, 1950-keV, and 1020-keV $\gamma$ rays [Figs.~\ref{fig:fig8}(c) and (d)]. The energy summations for each possible decay path stemming from the 4.437-MeV level were in agreement to within $\sim1$~keV. Therefore, within the $\sim$1~keV energy uncertainty a single level at 4.437 MeV has been placed in the level scheme having four outgoing transitions (760-keV, 780-keV, 1609-keV, and 438-keV). A possible fifth transition through the 822-keV $\gamma$-ray was also included based on energy arguments alone. The 1577-keV and 1020-keV gated recoil-$\gamma$-$\gamma$ coincidence projections, both of which were found to feed the 4.437-MeV level directly, showed coincidences with the 760-keV, 780-keV, 1609-keV and 438-keV $\gamma$ rays. Unfortunately, a check for consistency between relative yields for the exiting transitions was not possible due to the low statistics and background counts in the $\gamma$-ray coincidence data. The 1950-keV and 1625-keV lines, and to a lesser extent the 559-keV line, each had a coincidence spectrum which supported a single level at 4.437-MeV [Figs.~\ref{fig:fig8}(e) and (f)]. 

The placement of only a single level at 4.437-MeV is in contrast to the proposed levels in Table 6.15 of Ref.~\cite{ref:Wang} but supported by the coincidence data and within the uncertainties of that work. The relative intensities between the 438-keV and 1609-keV lines are consistent between the two works. The 4.437-MeV level has been given a tentative range of $J=(4,5)$ with the $J=(5)$ assignment slightly favored due to the reaction mechanism and multipolarity information. From the linking transitions between the 4.437-MeV level to levels with known $J^{\pi}$, spins of $J=(4-6)$ were possible. The 760-keV transition placement was critical as it set the lower limit on $J$. The 1609-keV $R_{149^{\circ}/90^{\circ}}=0.74(13)$ ratio favors a dipole multipolarity and therefore, $J<6$. The 760-keV transition has a suggestive $R_{149^{\circ}/90^{\circ}}=0.72(35)$ ratio of a dipole transition and is consistent with values of $(5)\rightarrow 6^{(+)}$ for the 4.437-MeV and 3.677-MeV levels. The observed dipole multipolarity of the 780-keV transition was also consistent with either spin due to the uncertain final state spin of the 3.657-MeV level, $J=(3,4)$. The 438-keV and 822-keV multipolarities were inconclusive. It is unclear whether the 4.437-MeV level is the same as that observed in the ($t$,$p$) work of Ref.~\cite{ref:Dav85}. If it is, then a $J=(4)$, $\pi=+$ assignment could be made as the ($t$,$p$) proton angular distributions rule out $L=5$ and the 760-keV transition rules out a $J^{\pi}=3^-$ assignment.

\subsubsection{\label{subsec:1950}The 5.456-MeV, 6.014-MeV, 7.081-MeV, and 7.963-MeV levels}
Levels with energies of 5.456 MeV, 6.014 MeV, 7.081 MeV, and 7.963 MeV were identified in the present data. Only the 6.014-MeV level had parallels with some previous work which found levels at 6.020(30), 6.000(30), and 6.006 MeV~\cite{ref:May84,ref:Dav85,ref:Duf86}. It is likely that a common level having $J=(3^-)$ was observed in both the ($t,p$) and $\beta$ decay works at 6.006 MeV. However, this does not appear to be the same level observed in the present work due to disagreements in possible $J$ values. The strongest of the transitions comprising these new levels, the 1577-keV $\gamma$ ray, was first observed in the deep inelastic work of Ref.~\cite{ref:Wang2010} but here placed higher in the level scheme than postulated in their Table 6.15.

The combination of recoil-$\gamma$-$\gamma$ coincidence data with the 3.999-MeV and 4.437-MeV levels, as well as energy summations, led to the placement of the eight transitions (one as tentative) to or from these new levels. A lack of presence of the 438-keV $\gamma$ ray in the 1457-keV gated $\gamma$-ray spectrum and the relation between the 4.437-MeV level and the 1020-keV transition, established the 5.456-MeV level. The appearance of the 1625-keV transition in both the 1457-keV and 1020-keV gated spectra, and vice-versa, determined its location directly feeding the 5.456-MeV level. The 1950-keV to 1577-keV sequence was established through commensurate coincidences with one another as well as their observed intensities from the 1609-keV gated coincidence spectrum [Fig.~\ref{fig:fig8}(c)]. The relatively weak 559-keV transition linking the 6.014-MeV and 5.456-MeV levels was placed primarily based on its energy agreement, though it did also show tentative evidence for a coincidence relation with the 1950-keV [Fig.~\ref{fig:fig8}(f)] and 1020-keV transitions . Finally, the 1067-keV $\gamma$ ray was observed to be at least a doublet. One 1067-keV transition was in coincidence with the 1577-keV transition and its energy was consistent with the difference between the 7.082-MeV to 6.014-MeV levels. However, the placement of any further 1067-keV $\gamma$ rays was not able to be reconciled in the present work.

The 5.456-MeV level has been tentatively assigned $J=(5,6)$ based on the extracted dipole nature of the 1020-keV transition ($R_{149^{\circ}/90^{\circ}}=0.86(11)$) and the quadrupole nature of the 1457-keV transition ($R_{149^{\circ}/90^{\circ}}=2.16(67)$) which feed the $J=(4,5)$ 4.437-MeV and $J=(3,4)$ 3.999-MeV levels, respectively. The extracted multipolarities of the 1577-keV (quadrupole) and 559-keV (dipole) transitions similarly limit the $J$ range of the 6.014-MeV level to $J=(6,7)$. The 1950-keV transitions $R_{149^{\circ}/90^{\circ}}=1.52(40)$ ratio is consistent with a quadruple transition building upon the 6.014-MeV level giving $J=(8,9)$ for the 7.963-MeV level. As noted above in sub-section~\ref{subsec:yrast}, the suggested $J$ values for the 7.963-MeV level are consistent with the multipolarity of the tentatively placed 1617-keV transition which feeds into the $J=8^{(+)}_1$ 6.346-MeV level. Finally, only a limit of $J=6-8$ could be surmised for the 7.081-MeV level due to the doublet nature of the 1067-keV transition and the non-distinct $R_{149^{\circ}/90^{\circ}}$ value of the 1625-keV transition.

\subsubsection{Levels residing above 9.5~MeV in excitation energy}
Two additional levels at 9.885 MeV and 10.996 MeV have been placed into the $^{38}$S level scheme based on the observed 2804-keV and 3033-keV $\gamma$ rays, respectively. Due to the weak nature of these higher lying states, their placement was most apparent in the $\gamma$-ray coincidence data provided by the $k_{ml}>0.6$ recoil selection, though the 3033-keV line is visible in Fig.~\ref{fig:fig8}(e). The 2804-keV transition was within a few keV of the known 2.806-MeV $2^+_2$ level. However, it showed clear coincidences with the 1293-keV and 1535-keV transitions amongst others, leading to its placing higher in the level scheme. Furthermore, no evidence for a $\sim$2806-keV transition appeared in the 810-keV gated coincidence data which clearly showed the 1513-keV line. The 2804-keV line did appear in the summed $\gamma$-ray spectrum consisting of the combined 559-, 1020-, 1067-, and 1625-keV gated transitions and the 2804-keV coincidence gated spectrum reciprocated the stronger three of these transitions as well. The 2804-keV line also showed a preference for a dipole multipolarity, however, no $J$ information could be concluded for the 9.885-MeV level as only range of spins was placed on the 7.081-MeV level.

It is unlikely that the presently observed 3033-keV transition is the same as the 3010-keV line observed in the deep inelastic work~\cite{ref:Wang2010,ref:Wang} due to the energy differences. A clear coincidence was observed between the 3033-keV line and the 1293-keV transition which eliminated the possibility for direct ground-state feeding. The summed $\gamma$-ray coincidence spectrum comprised of both 1577-keV and 1950-keV gates demonstrated a clear coincidence with the 3033-keV line. A limit of $J\geq8$ could be made for the 10.996-MeV level based on the reaction mechanism and the tentative assignment of the 7.963-MeV level.

\subsubsection{Unplaced $\gamma$-ray transitions}
There were a sub-set of $\gamma$-ray transitions which were identified as belonging to the $^{38}$S level scheme but were unable to be placed. They have been labeled with parenthesis around their energies in Table~\ref{tab:tab2}. Of these, it is probable that the 2057-keV, 2195-keV, 2486-keV, and 2572-keV, correspond to transitions which have been previously listed in the deep inelastic work of Ref.~\cite{ref:Wang2010,ref:Wang}. It may be speculated that the 2057-keV transition feeds the 1.293-MeV level and that the 2195-keV level feeds the $J=4^+$ 2.827-MeV state, corresponding to the 3.375(17)-MeV and 5.064(27)-MeV levels observed in Ref.~\cite{ref:Dav85}. However, no coincidence data was available to weigh in on these hypotheses. 

\begin{table*}[t]
\caption{\label{tab:tab2}%
Identified $\gamma$-ray transitions in $^{38}$S based on the present work. Those which were unable to be placed in the level scheme are listed in parentheses. The $\gamma$-ray energies, E$_{\gamma}$, have uncertainties of $\sim$0.5-1~keV. The relative intensities have been normalized to the 1293-keV transitions ($\equiv1000$). The ratio of the angular yields, $R_{149^{\circ}/90^{\circ}}$, deduced multipolarities ($L$), and the relative intensities, $I_{\gamma}$, were extracted from data based on the $k_{ml}>0.60$ cut and include statistical and systematic uncertainties.}
\begin{ruledtabular}
\begin{tabular}{lccccccc}
\textrm{E$_{i}$ [MeV]} & $J^{\pi}_{i}$ & \textrm{E$_{\gamma}$ [keV]}
& \textrm{I$_{\gamma}$} & \textrm{E$_{f}$ [MeV]} & $J^{\pi}_{f}$ &
$R_{149^{\circ}/90^{\circ}}$ & Multipolarity ($L$) \\
\colrule
0.000	&	$0^+_1$	&	--	&	--	&	--	&	--	&	--	& --\\
1.293	&	$2^+_1$	&	1292.7	&	$\equiv1000$	&	0.000	&	$0^+_1$	&	1.29(5)	& 2\\
2.806	&	$2^+_2$	&	1513.1	&	38(6)	&	1.293	&	$2^+_1$	&	0.94(30)	& --\\
2.827	&	$4^+_1$	&	1534.5	&	598(35)	&	1.293	&	$2^+_1$	&	1.22(5)	& 2\\
3.520	&	(1-3)	&	2224.4	&	19(5)	&	1.293	&	$2^+_1$	&	--	& --\\
	&		&	3522.3\footnote{Placement is tentative.\label{a}}	&	24(6)	&	0.000	&	$0^+_1$	&	--	& --\\
3.615	&	(2,3)	&	788.3$^{\textrm{\ref{a}}}$	&	11(3)	&	2.827	&	$4^+_1$	&	0.76(48)	& --\\
	&		&	809.6	&	14(3)	&	2.806	&	$2^+_2$	&	0.64(29)	& --\\
	&		&	2322.8	&	140(12)	&	1.293	&	$2^+_1$	&	0.83(11)	& 1\\
3.657	&	(3,4)	&	830.2	&	37(4)	&	2.827	&	$4^+_1$	&	1.88(44)	& --\\
	&		&	2364.8	&	89(9)	&	1.293	&	$2^+_1$	&	0.99(19)	& --\\
3.677	&	$6^{(+)}_1$	&	850.1	&	201(12)	&	2.827	&	$4^+_1$	&	1.38(9)	& 2\\
3.999	&	(3,4)	&	383.6	&	39(5)	&	3.615	&	(2,3)	&	0.75(13)	& 1\\
4.437	&	(4,5)	&	438.3	&	18(3)	&	3.999	&	(3,4)	&	1.10(36)	& --\\
	&		&	759.7	&	16(2)	&	3.677	&	$6^{(+)}_1$	&	0.72(35)	& --\\
	&		&	779.8	&	40(4)	&	3.657	&	(3,4)	&	0.81(18)	& 1\\
	&		&	822.0$^{\textrm{\ref{a}}}$	&	15(4)	&	3.615	&	(2,3)	&	--	& --\\
	&		&	1609.3	&	86(7)	&	2.827	&	$4^+_1$	&	0.74(13)	& 1\\
5.456	&	(5,6)	&	1019.6	&	81(9)	&	4.437	&	(4,5)	&	0.86(11)	& 1\\
	&		&	1457.0	&	28(6)	&	3.999	&	(3,4)	&	2.16(67)	& 2\\
6.014	&	(6,7)	&	558.9	&	18(4)	&	5.456	&	(5,6)	&	0.64(21)	& 1\\
	&		&	1576.7	&	112(6)	&	4.437	&	(4,5)	&	1.25(16)	& 2\\
6.346	&	$8^{(+)}_1$	&	2668.2	&	194(25)	&	3.677	&	$6^{(+)}_1$	&	1.74(19)	& 2\\
7.081	&	(6-8)	&	1066.9\footnote{Identified as a doublet.}	&	69(6)	&	6.014	&	(6,7)	&	--	& --\\
	&		&	1625.4	&	45(5)	&	5.456	&	(5,6)	&	1.24(41)	& --\\
7.963	& (8,9)	&	1617.3$^{\textrm{\ref{a}}}$	&	40(5)	&	6.346	&	$8^{(+)}_1$	&	0.79(20)	& 1\\
	&		&	1950.3	&	48(6)	&	6.014	&	(6,7)	&	1.52(40)	& 2\\
8.730$^{\textrm{\ref{a}}}$	&	--	&	2384.7$^{\textrm{\ref{a}}}$ &	43(7)	&	6.346	&	$8^{(+)}_1$	&	1.61(73)	& --\\
9.885	&	--	&	2804.0	&	40(8)	&	7.081	&	(6-8)	&	0.70(28)	& 1\\
10.996	&	--	&	3032.6	&	17(7)	&	7.963	&	(8,9)	&	--	& --\\
--	&	--	&	(887)	&	21(5)	&	--	&	--	&	--	& --\\
--	&	--	&	(2015)	&	17(5)	&	--	&	--	&	--	& --\\
--	&	--	&	(2057)	&	34(6)	&	--	&	--	&	--	& --\\
--	&	--	&	(2195)	&	26(6)	&	--	&	--	&	--	& --\\
--	&	--	&	(2486)	&	25(6)	&	--	&	--	&	--	& --\\
--	&	--	&	(2572)	&	34(10)	&	--	&	--	&	--	& --\\
--	&	--	&	(3630)	&	14(5)	&	--	&	--	&	--	& --\\
--	&	--	&	(3764)	&	13(4)	&	--	&	--	&	--	& --\\
\end{tabular}
\end{ruledtabular}
\end{table*}

\section{\label{sec:dis}Calculations \& Discussion}

The low-lying positive-parity energy levels of $^{38}$S may be described by the coupling of a pair of protons occupying the $1s_{1/2}$ and $0d_{3/2}$ orbitals (outside of a filled $0d_{5/2}$ orbital) and a pair of neutrons which occupy the $0f_{7/2}$ and $1p_{3/2}$ orbitals. Within this space, $J$ values range from $0^+$--$8^+$ through the combination of $\pi(1s_{1/2}0d_{3/2})_{J=1,2}$, $\pi(1s_{1/2})^{2}_{J=0}$ or $\pi(0d_{3/2})^{2}_{J=0,2}$ configurations with $\nu(0f_{7/2}1p_{3/2})_{J=2-5}$, $\nu(0f_{7/2})^{2}_{J=0,2,4,6}$, or  $\nu(1p_{3/2})^{2}_{J=0,2}$ configurations. Excited levels reaching $J^{\pi}>8^+$ are most readily accessible from either the promotion of a $0d_{5/2}$ proton into the $1s_{1/2}0d_{3/2}$ orbitals or from the promotion of pairs of $1s0d$ particles into the upper $0f1p$ shell, so-called $2p-2h$ excitations. Though contained within the valence space of previous calculations, i.e. the SDPF-MU~\cite{ref:Uts01,ref:Lun16} and the SDPF-U~\cite{ref:Now09,ref:Wang2010} interactions, there were no published predictions of the locations of $J^{\pi}>6^+$ levels based on a $\pi(0d_{5/2})^{-1}$ configuration or high-$J$ states based on $2p-2h$ configurations as they were not the focus of those works. However, lower-$J$ values having $2p-2h$ character have been predicted to appear at around 3--4~MeV in excitation energy~\cite{ref:Dav85,ref:Oln86}. Experimentally, there are a few possible candidates for such $2p-2h$ states, for instance levels with natural parity that were weakly populated in previous $(t,p)$ or 2-$n$ transfer work~\cite{ref:Fif84,ref:May84,ref:Dav85,ref:Oln86,ref:War87}, but there are no concrete correspondences.

The lowest-lying negative parity states are expected to have single-particle configurations built upon odd numbers of particle-hole excitations across the $1s0d$-$0f1p$ shell gap, so-called $1p-1h$ excitations. Experimentally, a number of candidates exist for levels with negative parity based on the analysis of past $\beta$-decay and $(t,p)$ data~\cite{ref:Dav85,ref:Oln86,ref:War87,ref:Duf86}. The lowest energy candidate is around $\sim$3.5~MeV [$J^{\pi} = (1^- - 3^-)$], with other candidates residing in the $\sim$4 - 5~MeV region. The lowest candidate is well below both the $\sim4.5$--5~MeV predictions from the $\nu(0d_{3/2}^{4-n}0f_{7/2}^{n+2})^6$, $n=$~odd model-space constrained shell-model calculations presented in Ref.~\cite{ref:Oln86} as well as the weak-coupling prediction ($^{35}$S$\otimes^{43}$Ca) in Ref.~\cite{ref:Dav85}.

\subsection{Shell-model calculations based on the FSU interaction}
To further interpret the level scheme of $^{38}$S, shell-model calculations were completed using an empirically-adjusted effective interaction, the so-called FSU~\cite{ref:Lubna2020} interaction. This $0p$-$1s0d$-$0f1p$ shell interaction was developed to prosper in the vicinity of $^{38}$S by including a valence space for particles within both the $1s0d$ and $0f1p$ shells. There was an emphasis on describing intruder states consisting of particle-hole excitations ($np-nh$) across either the traditional $0p$-$1s0d$ shell closure or from within the $1s0d$ shell across the traditional $N=Z=20$ shell closure into the adjacent $0f1p$ sub-shells~\cite{ref:Lubna2020}. The FSU interaction has been successful in describing, amongst other things, levels based on the $\pi1s0d$- and $\nu0f1p$-shell configurations~\cite{ref:Chrisman2021,ref:MacGregor2021,ref:Williams2020} and in particular both normal and intruder levels in $^{38}$Cl and $^{38,39}$Ar~\cite{ref:Lub19,ref:Abromeit2019}.

\begin{figure*}[htp]
    \centering
    \includegraphics[width=0.9\textwidth]{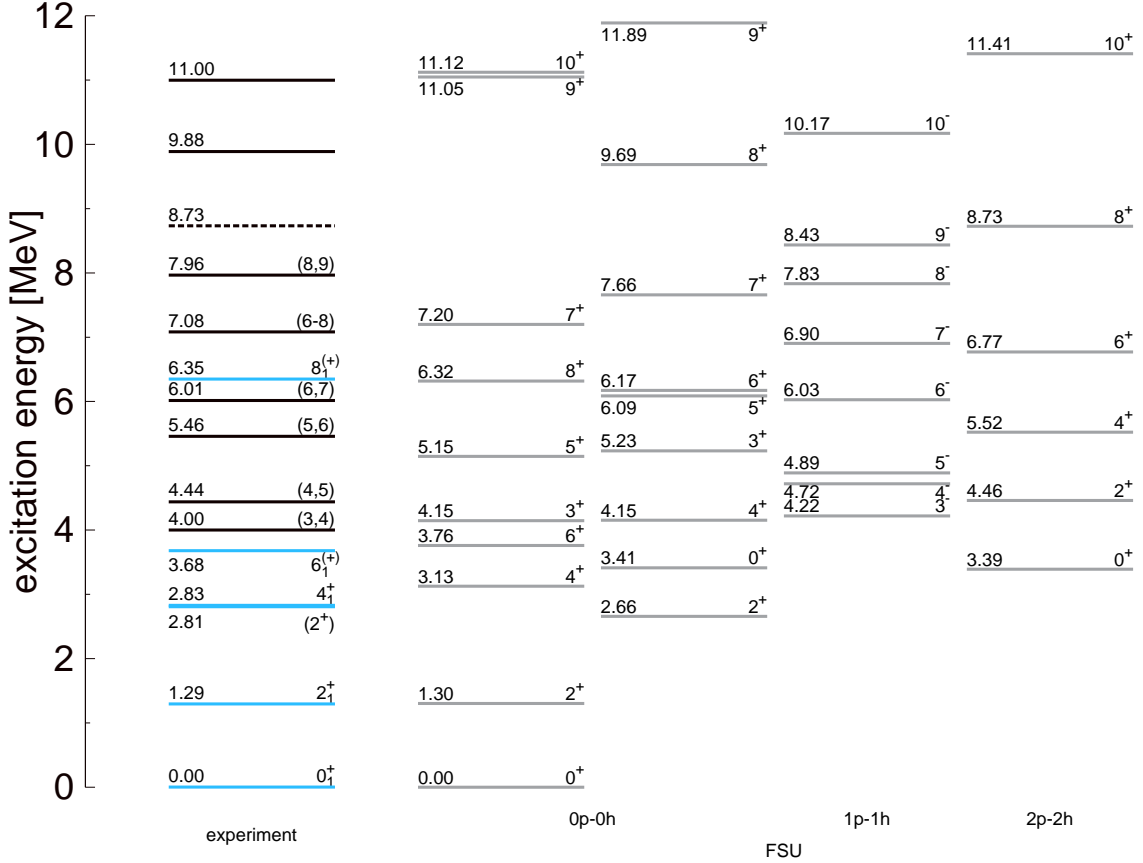}\\
    \caption{A subset of the experimental excited-state levels in $^{38}$S are presented along with $J^{\pi}$ assignments where available. Those in blue emphasize the established yrast even-$J$ levels. Calculated levels based on the FSU~\cite{ref:Lubna2020} interaction in which the protons were confined to the $1s0d$ shell and neutrons to the $1p0f$ shell are labeled as $0p-0h$. Those that allowed for one or two particle-hole excitations across the $N=20$ shell gap, from within the $1s0d$ shell to the $0f1p$ shell, are labelled by $1p-1h$ or $2p-2h$, respectively. Only a subset of the calculated levels are also shown (see text for additional details).} 
    \label{fig:fig9}
\end{figure*}

Positive parity $0p-0h$ states in $^{38}$S were calculated using the FSU interaction and requiring 8 valance protons to be confined within the $1s0d$ shell and the 2 valence neutrons to be confined within the full $0f1p$ shell. The interaction was also used to provide predictions of negative parity ($1p-1h$) states and additional positive parity ($2p-2h$) states by allowing fixed-numbers of excitations (protons or neutrons) to proceed across either the $N=Z=8$ $0p$-$1s0d$ shell gap or the $N=Z=20$ $1s0d$-$0f1p$ gap. No mixing was allowed between the positive parity $0p-0h$ and $2p-2h$ states. A subset of the calculated levels are presented in Fig.~\ref{fig:fig9} including the two lowest-energy $0p-0h$ states for spins through $J^{\pi}=10^+$. The lowest energy level for each spin up to $J^-=10^-$ is also shown for the $1p-1h$ calculations, and similarly, for even-$J^+$ states from the $2p-2h$ calculations.

The calculations carried out with the FSU interaction give similar low-lying level orderings and energies as the calculations presented in Refs.~\cite{ref:Dav85,ref:Oln86,ref:Lun16,ref:Stu06,ref:Longfellow2021}. The yrast $8^+$ level is calculated to have a $0p-0h$ configuration and reside at an energy of 6.32~MeV. The predicted lowest-energy positive parity $2p-2h$ excitations appear at $\sim$3~MeV in excitation energy in Fig.~\ref{fig:fig9}. However, the $2p-2h$ even-$J$ states do not begin to compete in energy with the $0p-0h$ levels so as to become yrast states up to at least $J^{\pi}=10^+$. There the lowest-lying $0p-0h$ 10$^+$ level is predicated at around 11.1~MeV, while the partner $2p-2h$ 10$^+$ energy is at 11.4~MeV, only 300~keV higher. The FSU interaction also predicts the lowest-energy negative parity states of $1p-1h$ character ($J^{\pi}=3^- - 5^-$) in the 4.2 -- 4.9~MeV energy region. This is consistent with the previous model predictions discussed above. The negative parity $1p-1h$ states are predicted to become yrast in spin at around $J=9$ or 10, and near 8.5~MeV in excitation energy, $\sim$1~MeV below the predicted $0p-0h$ 9$^+$ and 10$^+$ levels.

\subsection{The even-$J$ yrast states}
\begin{figure}[ht]
    \centering
    \includegraphics[width=0.48\textwidth]{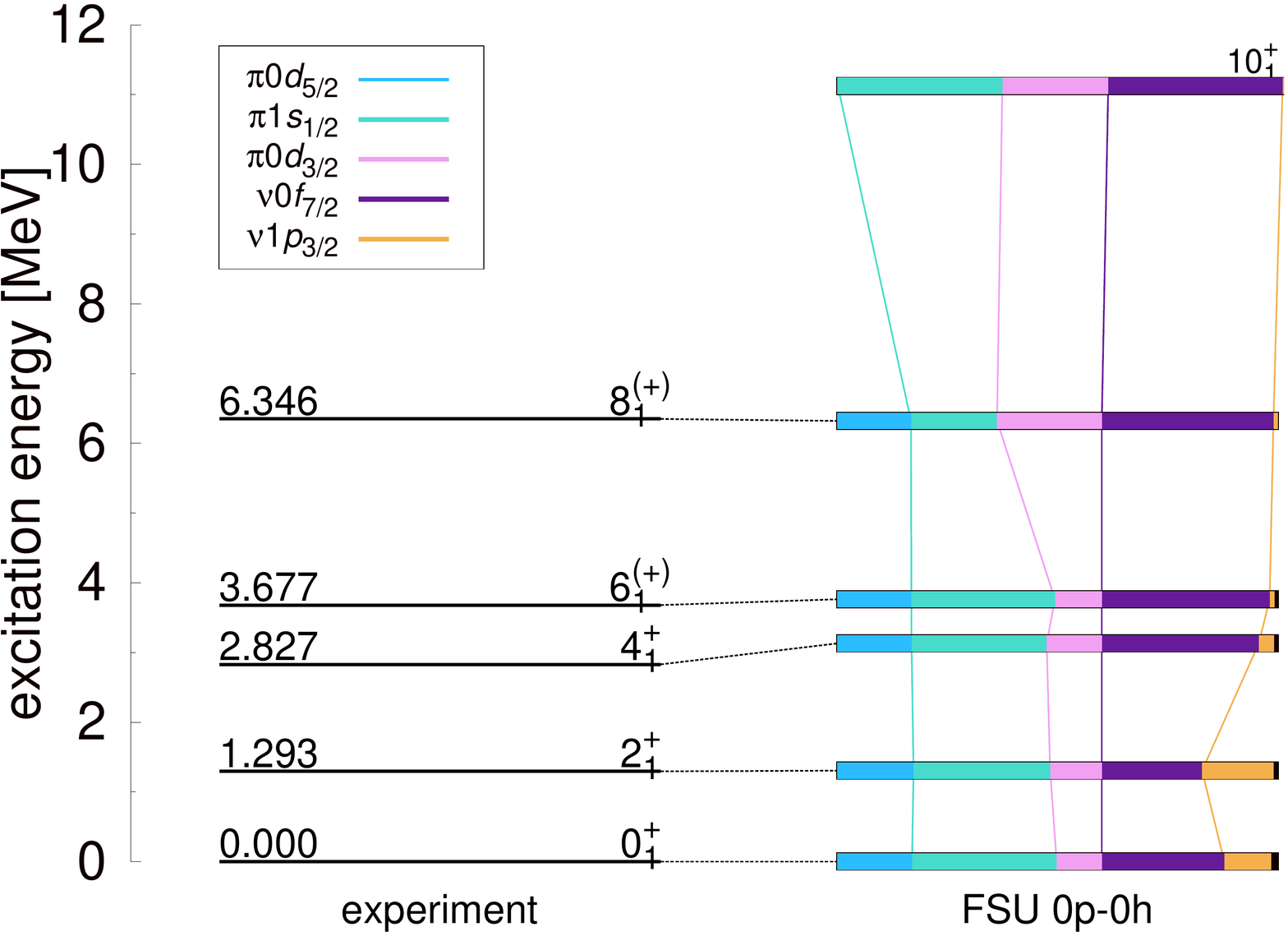}\\
    \caption{The excitation energies and calculated occupancies ($0p-0h$) of the even-$J^+$ yrast states in $^{38}$S through the $10^+$ level. The relative occupancies for the 'last' three $sd$-shell protons and the two neutrons within the $0f1p$-shell are labelled by color and connected by lines to highlight their evolution. The five remaining $sd$-shell protons fill the $\pi0d_{5/2}$ orbital and the occupancy is only for the `6th' proton, i.e. an occupancy of one on the plot represents a filled $\pi0d_{5/2}$ orbital ($0d_{5/2})^6$ and a value of zero gives $\pi(0d_{5/2})^{-1}$.}
    \label{fig:fig10}
\end{figure}

In addition to the excitation energies plotted in Fig.~\ref{fig:fig9}, the calculated energies and occupancies for the even-$J^{+}$ yrast levels up to $J^{\pi}=10^+$ ($0p-0h$) are also shown in Fig.~\ref{fig:fig10}. The calculations reproduce the excitation energies for the $J^{\pi}=2^+,4^+,6^+$ and $8^+$ levels, assuming positive parity for the latter two experimentally. The states with $J\leq6^{+}$ show mixed but near constant proton $1s_{1/2}$ and $0d_{3/2}$ occupancies while the neutron configurations evolve with increasing $J$ from mixed $0f_{7/2}-1p_{3/2}$ occupancy to a pure $(0f_{7/2})^2$ configuration. As discussed below and in Sec.~\ref{sec:p32}, the $\nu1p_{3/2}$ occupancy deviates from the simple $(1f_{7/2})^2_{J=0,2,4,6}$ seniority $\nu=2$ picture for the low-lying even-$J$ multiplet for these two levels. This deviation is a critical component in the microscopic description of the collective nature of these low-lying levels. At larger spin, however, the $\nu(0f_{7/2})^2$ configuration is restored out of the demand for increased angular momentum. Namely, in order to generate spins of $6^+-10^+$ within the $0p-0h$ space, a pure $\nu(0f_{7/2})^2$ neutron configuration is needed as the inclusion of any $1p_{3/2}$ neutrons is limited to $J_{\text{max}}=5\hbar$ as $(0f_{7/2}1p_{3/2})_{J=2-5}$.

In accordance with Fig.~\ref{fig:fig10} the predominant factor determining the relative energy spacing between the $6^+$ -- $8^+$ -- $10^{+}$ levels is the arrangement of the proton occupancies. Starting with $(0f_{7/2})^2_{J=6}$ from the neutrons, there is a migration in the proton occupancy from the $1s_{1/2}$ into the $0d_{3/2}$, gaining 2$\hbar$ in angular momentum, which extends $J^{\pi}=6^+$ up to $8^+$. The experimental spacing between the $J^{\pi}=6^+ - 8^+$ levels, $\Delta E_x=2.67$~MeV, is well reproduced by the FSU interaction, $\Delta E_x=2.56$~MeV. It can be concluded, unsurprisingly, that the FSU interaction has a solid grasp on the description of the proton $1s_{1/2}-0d_{3/2}$ single-particle energies and the associated matrix elements.

One method to construct a $J^{\pi}=10^+$ level is through the excitation of a proton from within the $\pi0d_{5/2}$ orbital into the $\pi1s_{1/2}$ orbital where the $(0d_{5/2})^{-1}$ proton hole provides the additional angular momentum. The FSU interaction predicts this configuration to be yrast at 11.122~MeV. A second option for generating $J^{\pi}>8^+$ levels is through $2p-2h$ configurations. The addition of two neutrons into the $0f_{7/2}$ and $1p_{3/2}$ orbitals increases the neutron $J_{\text{max}}$ contribution to $J_{\text{max}}>6$ via the $(0f_{7/2})^{4}_{J_{\text{max}}=8}$ and $(0f_{7/2}1p_{3/2})^4_{J_{\text{max}}=9}$ configurations. The lowest-lying $2p-2h$ $10^+$ state predicted by the FSU interaction is only $\sim300$~keV above its corresponding $0p-0h$ level. Hence, the calculations point towards a possible mixing between these two states experimentally. Unfortunately, no solid candidates were empirically determined for either $10^+$ level though the state at 10.996~MeV is in the approximate energy range of the predictions.

\subsection{The role of the $\nu1p_{3/2}$ occupancy\label{sec:p32}}
Appreciable occupancy of the neutron $1p_{3/2}$ orbital, and the implicit weakening of the traditional $N=28$ shell closure, is known to play a role in the low-lying structure of the neutron-rich S isotopes. Together with the neutron $0f_{7/2}$ orbital, these orbitals provide the foundation for the emergence of a region of low-lying deformation in the neutron-rich $N=28$ isotones~\cite{ref:Now09,ref:Utsuno2012} via coherent proton-neutron correlations. The inclusion and occupancy of the $\nu1p_{3/2}$ orbital specifically, has been required for a proper description of the experimental transition rates~\cite{ref:Lun16,ref:Longfellow2021} and the $2^+_1$ excited-state $g$ factor~\cite{ref:Dav06,ref:Stu06} in $^{38}$S. In the work of Ref.~\cite{ref:Stu06} the role of the $\nu1p_{3/2}$ orbital in dictating the proton occupancies of the $1s_{1/2}$ and $0d_{5/2}$ via coherent quadrupole correlations was emphasized. Such correlations and the migration of nucleon occupancy led to a correct description of the observed $2^+_1$ $g$ factor while also correctly increasing the magnitude of the calculated quadrupole transition strength [B($E2$)] of this state. 

One seemingly conflicting piece of data with the coherent proton-neutron picture is the extracted ratio of the neutron-to-proton multi-pole transition matrix elements, $M_n/M_p=(1.5\pm0.3)N/Z$, for the $2^+_1$ state~\cite{ref:Kel97}. The $\gtrsim 1$ value points towards a non-symmetric (isovector) contribution of neutrons-to-protons to the $2^+_1$ excitation typically indicative of a closed nucleon shell, in this case a closed proton shell. As discussed in Ref.~\cite{ref:Kel97}, similar behavior has been observed in $^{18}$O, and to a less degree in $^{42}$Ca. One possible reconciliation is that while the $\nu1p_{3/2}$ occupancy is crucial to driving coherent behavior there is still a large fraction of the $\nu(0f_{7/2})^2$ configuration within the $2^+_1$ and ground state wave functions which manifests itself in the individual muliti-pole transition matrix elements differently than in the $g$ factor value.

\begin{figure}[ht]
    \centering
    \includegraphics[width=0.48\textwidth]{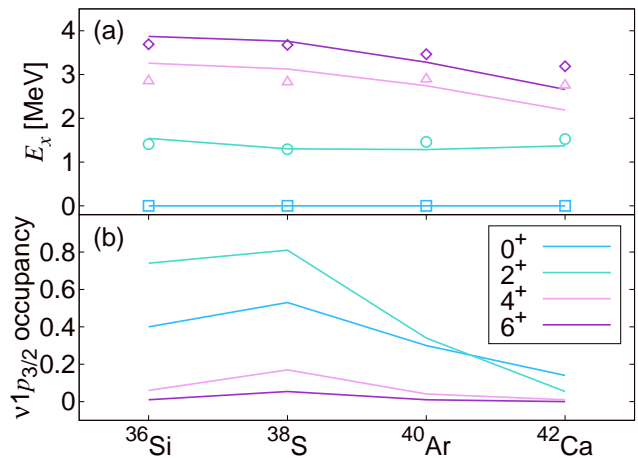}\\
    \caption{The even-$J^+$ yrast excitation energies ($E_x$) and $\nu$1p$_{3/2}$ occupancies from the $0p-0h$ shell-model calculations using the FSU interaction~\cite{ref:Lubna2020} (lines) for a selection of even $Z$ $N=22$ isotones. The experimental energies for the $J^{\pi}=0^+_1$ (blue squares), $2^+_1$ (green circles), $4^+_1$ (pink triangles), and $6^+_1$ (purple diamonds) are also given~\cite{ref:ensdf}.}
    \label{fig:fig11}
\end{figure}

The calculated excitation energies and neutron $1p_{3/2}$ occupancies from the FSU interaction ($0p-0h$) are shown by the lines in Figs.~\ref{fig:fig11}(a) and (b), respectively, for the Si, S, Ar, and Ca $N=22$ isotones. Both $^{36}$Si and $^{38}$S show distinct increases in their $1p_{3/2}$ occupancy ($\gtrsim 0.4$ nucleons for the $0^+_1$ and $2^+_1$ levels) relative to the heavier isotones of $^{40}$Ar and $^{42}$Ca. As discussed in the work of Ref.~\cite{ref:Liang2006}, and shown in Fig.~\ref{fig:fig11}, striking similarities exist experimentally between the low-lying even-$J$ levels of $^{36}$Si and $^{38}$S. The energies of their yrast levels through $J^{\pi} = 6^+$ and their ground state to $2^+_1$ dynamic quadrupole transitions strengths, B($E2,0^+_1 \rightarrow 2^+_1$) values, agree within $\sim$10\% of each other. The FSU interaction calculations reproduce the excitation energies well [Fig.~\ref{fig:fig11}(a)] and slightly under predict the B($E2,0^+_1 \rightarrow 2^+_1$) values but reproduce the trend. Additionaly, the $\nu1p_{3/2}$ occupancies from the FSU interaction are larger than those predicted by the SDPF-MU interaction for the same two levels in $^{38}$S by $>0.2$ nucleons (see Fig.~20 of Ref.~\cite{ref:Lun16}). This may be an indicator as to why there is an improper spacing between the $^{38}$S $0^+_1, 2^+_1$ energies relative to the $4^+_1,6^+_1$ level energies by about $\sim350-400$~keV (see Fig. 4 of Ref.~\cite{ref:Lun16}).

Interpretation of the above information and that in Fig.~\ref{fig:fig11} suggests similar magnitudes of low-lying deformation. Even still, the slight increase in the $1p_{3/2}$ occupancy ($\sim$0.1 nucleons) in $^{38}$S is consistent with it having a lower 2$^+_1$ energy and a larger B($E,0^+_1 \rightarrow 2^+_1$) value. It is perhaps initially surprising that the highest $1p_{3/2}$ occupancy does not reside in the most neutron-rich system, $^{36}$Si, considering its neutron $0f_{7/2}$-$1p_{3/2}$ energy spacing should be the most reduced. However, it appears that the correlation energy gained by the arrangement of the two additional protons in $^{38}$S is enough to overcome a change in a slight increase in the neutron orbital energy spacing. It should be noted, that while the magnitude of the deformations in $^{36}$Si and $^{38}$S are similar, it has been postulated theoretically that $^{36}$Si has an axially-symmetric oblate shape compared to a prolate shape in $^{38}$S~\cite{ref:Utsuno2012}. The prolate axially-symmetric shape in $^{38}$S has been effectively established by the $g$-factor work~\cite{ref:Dav06,ref:Stu06} while experimental information for $^{36}$Si is lacking. 

\subsection{The neutron $0f1p$ shell states}
The coupling of a single neutron in each the $0f_{7/2}$ and $1p_{3/2}$ orbitals results in a multiplet of states ranging from $J=2^{+}-5^{+}$. While the lower spins of this multiplet are susceptible to mixing, a correspondence between states with this ideal single-particle configuration was found with the calculations of the FSU interaction for the $5^+_1$ and $4^+_2$ levels at 5.146~MeV and 4.153~MeV, respectively. The calculated neutron occupancies for these two levels, $0f_{7/2}^{\sim1.2}$ and $1p_{3/2}^{\sim0.7}$, close to the single-particle picture, and the calculated proton occupancies were also similar, ($1s_{1/2}^{\sim1.5}$ and $0d_{3/2}^{\sim0.5}$). There is a predicted $3^+_1$ level by the FSU interaction at 4.146~MeV that is a possible member of the multiplet, though it shows variation in the proton occupancy and an increase of 0.2 in the $\nu0f_{7/2}$ occupancy relative to the aforementioned $4^+_2$ and $5^+_1$ levels. Under the assumption that the $5^+_1$ level specifically, has a reasonably pure $(0f_{7/2}1p_{3/2})$ neutron configuration and the $6^{+}_{1}$ level has a nearly pure $(\nu0f_{7/2})^2$ configuration (Figs.~\ref{fig:fig10} and ~\ref{fig:fig11}), the energy spacing between these levels is primarily dependent upon the description of the $0f_{7/2}$-$1p_{3/2}$ interaction and their single-particle energies. The calculated energy difference between the $5^+_1$ and $6^+_1$ states is $\Delta E_x \approx 1.4$~MeV. Experimentally, there are a series of levels at 3.615~MeV, 3.999~MeV, and 4.437~MeV, that have been eluded to as belonging to this $(0f_{7/2}1p_{3/2})_{J=2-5}$ multiplet~\cite{ref:Oll04,ref:Wang2010,ref:Wang}. Assuming this sequence of levels are the $J^{\pi}=2-4^+$ multiplet members, and the newly established level at 5.456~MeV is the $5^+$ candidate, a consistent (though non-unique) picture between theory and experiment emerges. Under this scenario, $\Delta E_x \approx 1.8$~MeV for the energy difference between the experimental $5^+_1$ and $6^+_1$ levels, which suggests some discrepancy with the calculations ($\approx400$~keV). However, as mentioned, there are a other plausible corresponding $J$ assignments between the observed 5.456~MeV level and calculation, including the $5^+_2$, $5^-_1$, $6^+_2$, and $6^-_1$ calculated levels.

\subsection{Other possible correspondences between the observed and calculated levels}
The measured state at 4.437 MeV has additional counterparts in the calculations beyond the suggested $J^{\pi}=4^+$ assignment and its membership in the $(0f_{7/2}^{1}1p_{3/2}^1)_{J=2-5}$ multiplet discussed above. In particular, the relatively large amount of fractional incoming and outgoing yield through the level may suggest it is a low-lying negative parity intruder state. The calculated $J^{\pi}=4^-$ and $5^-$ states are nearby at $\sim4.7-4.9$~MeV. Based on energy arguments alone, the subsequent 6.014-MeV and 7.963-MeV levels, each connected via quadrupole transitions (Fig.~\ref{fig:fig7} and Table~\ref{tab:tab2}) agree best with the $J^{\pi} = 6^-$ and $8^-$ levels at 6.027 and 7.832~MeV. Though a lowering of the calculated $J^{-}$ states on the order of $\sim$300-400~keV, would also create a viable scenario for the $9^- \rightarrow 7^- \rightarrow 5^-$ sequence. 

\section{\label{sec:sum}Summary \& Conclusions}

The level scheme of $^{38}$S has been extended in both excitation energy ($\sim$11~MeV) and maximum spin ($J\gtrsim8$) based on new in-beam $\gamma$-ray data. $^{38}$S was populated through a fusion-evaporation reaction involving a beam of $^{22}$Ne interacting with an $^{18}$O target at the Argonne National Laboratory ATLAS Facility. Prompt $\gamma$ rays emanating from the reaction were detected by the GRETINA array and an event-by-event selection of the recoiling nuclei was carried out by the Fragment Mass Analyzer. In order to improve upon the traditional methods used in the unique recoil identification of $^{38}$S, machine-learning techniques were employed for the first time. The supervised training of a feed-forward neural network in a One-Vs-All mode was carried out using labelled data based on known $\gamma$-ray energies and standard training procedures. Clear signatures were observed in the distribution and behaviour of the output values from the fully-trained model for $\gamma$-ray transitions belonging to $^{38}$S relative to those belonging to other isotopes or backgrounds. Hence, the selection of specific ranges of model output values provided a far-improved determination of $\gamma$-ray transitions in $^{38}$S utilizing both singles and coincidence $\gamma$-ray spectra.

The extension of the known even-$J$ yrast levels up through $J^{\pi}=8^{(+)}$ facilitated a discussion on their energy spacing and underlying single-particle structures. In particular, the energy spacing between the $6^{(+)}_1$ and $8^{(+)}_1$ levels was used to extract the amount of energy required to promote a proton from the $1s_{1/2}$ orbital into the $\pi0d_{3/2}$ orbital, $\Delta E_x = (6.346~\mathrm{MeV} - 3.677~\mathrm{MeV}) \approx 2.7$~MeV. Furthermore, candidates for the high-$J$ levels belonging to the near-pure neutron $(0f_{7/2}1p_{3/2})_{J=2 - 5}$ configuration were identified, for instance the level at 5.465~MeV was postulated as the $5^+$ member. The energy difference between this $5^+_1$ level and that aforementioned $6^+_1$ yrast state, $\Delta E_x = (5.465~\mathrm{MeV} - 3.677~\mathrm{MeV}) \approx 1.8$~MeV, is determined by the rearrangement of the $0f_{7/2}$-$1p_{3/2}$ occupancies, their corresponding single-particle energies, and their interactions.

Shell-model calculations incorporating the FSU interaction~\cite{ref:Lubna2020} well reproduced the energies of the even-$J$ yrast sequence and low-lying level scheme in general. They also provided guidance on a number of possible spin-parity scenarios for other corresponding states. The calculations reproduced the measured energy spacing between the $6^+_1$ -- $8^+_1$ levels giving to the interactions accurate description of the proton single-particle energies and interaction strengths within the $1s0d$ shell. A discrepancy of $\sim$400~keV was noted between the theoretical spacing of the $6^+_1 - 5^+_1$ levels and the experimental value, noting however that the experimental $5^+_1$ assignment was speculative. Furthermore, the calculations supported discussions pertaining to the spectroscopic similarities between $^{36}$Si and $^{38}$S. In particular, the striking resemblance of their neutron $1p_{3/2}$ occupancy across the even-$J^+$ yrast levels~\cite{ref:Liang2006}, excitation energies, and B($E2,0^+_1 \rightarrow 2^+_1$) values. The role of the neutron $1p_{3/2}$ occupancy and its importance in developing coherent proton-neutron correlations in the low-lying states of the $N=22$ systems was also reiterated, building upon the previous discussions in Refs.~\cite{ref:Dav06,ref:Stu06,ref:Utsuno2012,ref:Now09,ref:Lun16,ref:Longfellow2021}. In closing, a few open questions about $^{38}$S stemming from the present work include: i) why is the $(M_n/M_p)$ value extracted for the $2^+_1 \rightarrow 0^+_1$ transition from inelastic proton scattering data not consistent with the hydrodynamical, $N/Z$, limit? ii) where is the location of the yrast $10^+$ level and how mixed is this state? iii) where do the negative parity intruder levels appear for certain and where do they become yrast?

\section{Acknowledgments}
This material is based upon work supported by the U.S. Department of Energy, Office of Science, Office of Nuclear Physics, under Contracts No. DE-AC02-06CH11357 (Argonne), No. DE-FG02-94ER40848 (UML), No. DE-AC02-05CH11231 (LBNL), and No DE-SC0020451 (FRIB). This research used resources of ANL’s ATLAS facility, which is a DOE Office of Science User Facility. We gratefully acknowledge the computing resources provided on Bebop, a high-performance computing cluster operated by the Laboratory Computing Resource Center at Argonne National Laboratory. GRETINA was funded by the U.S. DOE, Office of Science, Office of Nuclear Physics, and operated by the ANL and LBNL contract numbers above. TRIUMF receives federal funding via a contribution agreement through the National Research Council Canada (NRC). Work at LLNL was performed under DOE Contract No. DE-AC52-07NA27344 and was supported by the LLNL-LDRD Program under Project No. 23-LW-047. Shell model calculations used the computational facilities of Florida State University supported by the Grant No. DE-SC0009883 (FSU).

\bibliography{references}%

\end{document}